\documentclass[%
 reprint,
 amsmath,amssymb,
 aps,
 pra
]{revtex4-1}

\usepackage{graphicx}
\usepackage{subfigure}
\usepackage{dcolumn}
\usepackage{bm}
\usepackage{color}

\newcommand{\dd}{\text{d}}
\newcommand{\Tr}{\text{Tr}}

\newcommand{\ket}[1]{\left| #1 \right\rangle}

\newcommand{\expectation}[1]{\left\langle #1 \right\rangle}
\newcommand{\modulus}[1]{\left| #1 \right|}

\newcommand{\commentout}[1]{}

\begin{document}


\title{Dissipation-Induced Mobility and Coherence in Frustrated Lattices}

\author{E. T. Owen}
 \email{e.owen@hw.ac.uk}
\author{O. T. Brown}
\author{M. J. Hartmann}
\affiliation{Institute of Photonics and Quantum Sciences, Heriot-Watt University Edinburgh EH14 4AS, United Kingdom}

\date{\today}

\begin{abstract}

In quantum lattice systems with geometric frustration, particles cannot move coherently due to destructive interference between tunnelling processes.  Here we show that purely local, Markovian dissipation can induce mobility and long-range first-order coherence in frustrated lattice systems that is entirely generated by incoherent processes.   Interactions reduce the coherences and mobility but do not destroy them.  These effects are observable in experimental implementations of driven-dissipative lattices with a flat band and non-uniform dissipation.

\end{abstract}


\maketitle

\section{Introduction} 

The wave functions of a perfect crystal are described by Bloch states which are delocalised over the entire crystal.  Transport through these states is possible as particles put into the system by a source can be extracted at a remote location by a drain.  However, interference can strongly affect the crystal's transport properties.  For example, disorder or impurities in the crystal create multiple scattering centers, leading to Anderson localisation~\cite{Anderson:1958, Lee:1985, Segev:2013}.  Destructive interference localises the wave functions which can no longer transport particles through the crystal.  Recently, the joint effects of interactions and interference resulting in many-body localization have attracted enormous interest~\cite{Nandkishore:2015, Altman:2015, Choi:2016, Smith:2016, Luitz:2016}. Also in the context of biological systems, the effect of dephasing on interference in transport processes has recently been studied for light harvesting complexes~\cite{Plenio:2008, Mohseni:2008, Caruso:2009, Huelga:2013}.

Localisation can also exist in crystals without disorder where geometric constraints on the tunnelling rates lead to destructive interference. Frustration quenches the kinetic energy of the Bloch states resulting in a flat band with an infinite effective mass.  This macroscopic degeneracy allows localised wave functions to be constructed which are insulating stationary states of the system.  Synthetic flat band crystals have recently been created in photonic lattices~\cite{Guzman-Silva:2014, Vicencio:2015, Mukherjee:2015A, Mukherjee:2015B}, polaritons in etched semiconductor heterostructures~\cite{Jacqmin:2014, Baboux:2016}, ultracold atomic gases in optical lattices~\cite{Taie:2015} and surface plasmons~\cite{Nakata:2012, Kajiwara:2016}. Moreover, concepts for the realisation of such lattices have been put forward for superconducting resonators~\cite{Yang:2016}.  

In several recently considered samples the explored particles of the system are photons~\cite{Hartmann:2016}. Due to unavoidable experimental imperfections, these are usually prone to loss.  In order to repopulate the crystal, a constant coherent or incoherent drive can be applied.  Therefore, these driven-dissipative frustrated lattices provide a fascinating new platform with which to study localisation, coherence and mobility phenomena.

The first works that have recently started to explore geometrically frustrated lattices in such regimes investigated to what extent frustration phenomena of closed systems, such as the enhancement of interaction effects, survive in the presence of driving and dissipation~\cite{Biondi:2015, Casteels:2016}.  Whereas the effect of frustration and interactions has been extensively investigated~\cite{Huber:2010, Moller:2012, Nie:2013, Tovmasyan:2013, Takayoshi:2013, Phillips:2015, Pudleiner:2015}, the interplay of frustration and local dissipation is much less explored.  

Here we show that despite the quench of the kinetic energy in frustrated systems, localized particles are enabled to move by dissipation, even if this dissipation is strictly local, i.e. each lattice site couples to its own bath. We demonstrate that local Markovian dissipation in a crystal with a flat band permits the transfer of excitations between independent frustrated states and generates coherences between these states. Any mobility due to mixing of flat and dispersive bands can be neglected provided their energy separation is much larger than the system-environment couplings.  Whilst the lattice sites interact locally with their own baths, frustrated states share these baths such that second-order processes, whereby an excitation hops into the bath and back into the system, allow excitations to tunnel from one frustrated state to another and generate first order coherence between these states.  The generation of coherence by dissipation also distinguishes this scenario from pure dephasing which would preserve the number of excitations.  Moreover, the dissipation terms we consider here are different from dissipators that can be engineered via coherent driving~\cite{Diehl:2008} as they do not conserve the number of excitations.

We detail the model in Section~\ref{sec:Model} and show that local Markovian baths generate dissipative coupling terms between the flat band states of a frustrated lattice.  To explore dissipation-induced mobility we introduce a localised pump and investigate the excitation density in lattice sites away from the pump in the stationary state.  In Section~\ref{sec:NonInteracting}, we find an increased density of excitations that, in the absence of interactions, decays exponentially in the distance from the location of the pump.  We also investigate coherence between localised Wannier states and find that first order coherence is generated by the dissipation even for an incoherent input.  This coherence is long range but is reduced in the vicinity of incoherent pumps.  In Section~\ref{sec:MobilityModelling}, we show that these results cannot be solely attributed to direct coupling between adjacent sites and that long-range incoherent hopping processes need to be taken into account in order to accurately describe the mobility.  We then explore the effect of interactions in Section~\ref{sec:StronglyInteracting}, showing that dissipation-induced mobility and coherence is a generic effect for frustrated systems.  We state our conclusions in Section~\ref{sec:Conclusions} and discuss possible experimental platforms where these effects can be explored.

\section{Model}
\label{sec:Model}

\begin{figure}[t]
    \centering
    \includegraphics[width=\columnwidth]{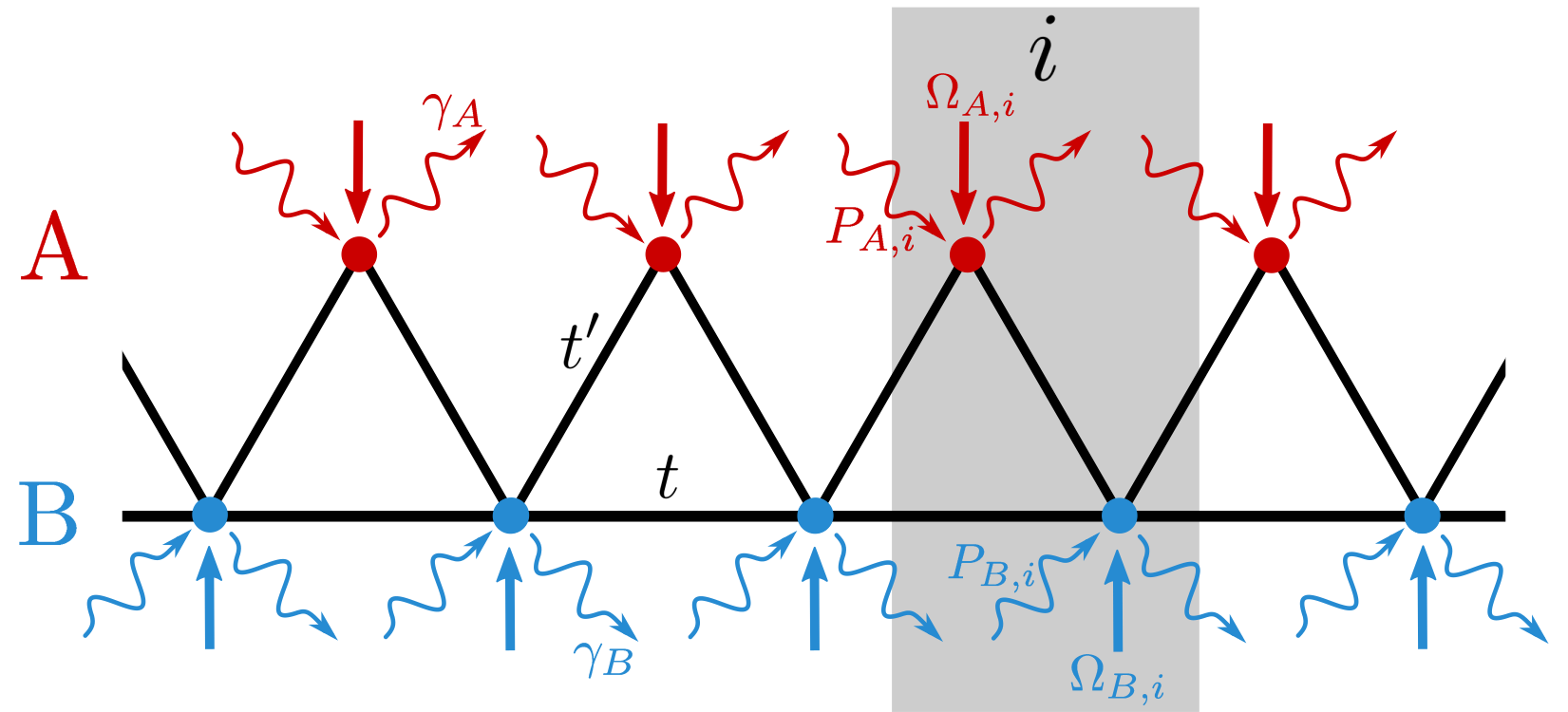}
    \caption{(Color online) The lattice with tunnelling rates $t$ and $t'$ along with the labelling the A and B sublattices and the unit cell $i$.  Excitations dissipate at a rate $\gamma_{A (B)}$ from the individual sites of the A (B) sublattice and coherent or incoherent drives are applied with amplitude $\Omega_{X, i}$ or intensity $P_{X, i}$ (X = A, B).
      }
    \label{fig:lattice_diagram}
\end{figure}

The sawtooth lattice consists of triangles connected to each other at the apex of each triangle, see Fig.~\ref{fig:lattice_diagram}.  The unit cell consists of two sublattices A and B where the A sites are not connected and excitations have a tunnel coupling $t$ between adjacent B sites and $t'$ between neighbouring A and B sites. 

The Hamiltonian $H = H_t + H_U + H_D$ of our lattice is the sum of a tight binding Hamiltonian $H_t$, an interaction Hamiltonian $H_U$ and a driving Hamiltonian $H_D$.
\begin{equation}
    \label{eq:H_t}
    H_t = H_0 + \sum_i \left(t b^\dag_{i-1} b_i + t' (b^\dag_i a_i + b^\dag_i a_{i-1}) + \mathrm{h.c.} \right)
\end{equation}
where $a_i\ (b_i)$ are the annihilation operators on site $i$ for the A (B) sublattice and $H_0 = \sum_i \omega_0 (a^\dag_i a_i + b_i^\dag b_i )$ with $\omega_0$ being onsite energy of each site (we set $\hbar = 1$ and all lengths are expressed in units of the lattice spacing).  The on-site interactions are of the form
\begin{equation}
    \label{eq:Interaction}
    H_U = \sum_i \left(U_A a^\dag_i a^\dag_i a_i a_i + U_B b^\dag_i b^\dag_i b_i b_i \right)
\end{equation}
where $U_{A (B)}$ is the interaction strength for multiple occupation of sites on the A (B) sublattice. The driving Hamiltonian will be specified in Eq.~(\ref{eq:H}) below.  We start by considering the non-interacting case $U_A = U_B = 0$ and return to investigate the influence of interactions later.

The Hamiltonian of Eq.~(\ref{eq:H_t}) can be written in terms of decoupled Bloch modes with frequencies $E_k = \omega_0 + t \cos k \pm \sqrt{t^2 \cos^2 k + 2 t'^2 (1 + \cos k)}$.  In the limit $t' \to \sqrt{2} t$, the lower-energy band of the Hamiltonian Eq.~(\ref{eq:H_t}) becomes flat with a gap of $2 t$ to the dispersive band.  The kinetic energy in the flat band is quenched due to geometric frustration. This is clearly seen by writing $H_{t}$ in a basis of Wannier states which are exponentially localised around the unit cell $i$.  The tight-binding bosonic operators can be expressed in terms of operators $W_i^\dag$ ($W_{i}$), that create (annihilate) excitations in the Wannier states, as
\begin{equation} 
  \label{eq:a_operator_def}
  \left. \begin{array}{r}
            a_i^\dag\\
            b_i^\dag
  \end{array} \right\} 
  = \sum_j \left\{ \begin{array}{c}
             w_A(r_i - r_j)\\
             w_B(r_i - r_j)
           \end{array}
  \right\} W_j^\dag + \mathrm{dispersive\ band}
\end{equation}
where $w_A (r) = \frac{\sqrt{2}}{2 \pi} \int_{-\pi}^{\pi} \dd k \cos (k / 2) e^{- i k r} e^{i k / 2} /\sqrt{\cos k + 2} $ and $w_B (r) = \frac{- 1}{2 \pi} \int_{-\pi}^{\pi} \dd k e^{-i k r}/\sqrt{\cos k + 2}$ are independent of $t$ due to the flat-band criterion.  If the band gap is sufficiently large, excitations into the dispersive band are negligible and we can project the Hamiltonian onto the flat band~\cite{Huber:2010} such that $H_t = (\omega_0 - 2 t) \sum_i W_i^\dag W_i$, see App~\ref{app:MasterEquationDerivation} for details.  Here,  one clearly sees that the localised Wannier states decouple and there is no propagation in $H_t$.

To explore mobility and coherence in this driven-dissipative regime, we first consider a scenario where one Wannier orbital is pumped by a coherent drive with amplitude $\Omega_W$ and frequency $\omega_D$.  In a frame rotating at $\omega_D$ and ignoring rapidly rotating terms, for $U_A = U_B = 0$ we get
\begin{align}
    \label{eq:H}
    H = \sum_i \Delta W_i^\dag W_i + \frac{1}{2} \left( \Omega_W W_0 + \Omega_W^* W_0^\dag \right)
\end{align}
where $\Delta = \omega_0 - 2 t - \omega_D$ is the detuning of the drive from the flat band and $i = 0$ is the pumped site.  We consider a drive on resonance with the flat band so $\Delta = 0$.  In an experiment, this drive of a Wannier state can be implemented by driving the A and B sites coherently with a frequency in resonance with the flat band and with suitable amplitudes such that $\Omega_W W_0 = \sum_j \Omega_{A, j} w_A (r_{i_0} - r_j) a_j + \Omega_{B, j} w_B (r_{i_0} - r_j) b_j$.  

The dissipation of excitations from the lattice is treated in the Born-Markov approximation with independent baths for each of the lattice sites.  The system is thus described by the master equation $\dot{\rho} = - i [H, \rho] + (1/2) \sum_i (\gamma_A D[a_i] \rho + \gamma_B D[b_i] \rho)$ where $\gamma_A\ (\gamma_B)$ is the dissipation rate for the A (B) sublattice, $D[x] \rho = 2 x \rho x^\dag - \{ x^\dag x, \rho\}$ is a Lindblad-type dissipator and $\{ \cdot, \cdot \}$ denotes the anti-commutator.  Expressing the dissipation in terms of the Wannier states, we get
\begin{equation}
    \label{eq:site_master_equation}
    \dot{\rho} = - i [H,\rho] + \frac{1}{2} \sum_{j, l} \gamma_{l} (2 W_j \rho W^\dag_{j+l} - \{W^\dag_{j+l} W_j, \rho \}) 
\end{equation}
with $\gamma_l = \sum_{i} \sum_{x=A,B} \gamma_x w_x(r_{i}) w_x (r_{i} - r_{l})$ leading to non-local dissipation coefficients
\begin{equation}
    \frac{\gamma_l}{\gamma_A} = \left\{ \begin{array}{ll} (2 f_0 + f_1) - (1-\kappa)f_0 & \mathrm{for} \ l = 0 \\
                                   - (1 - \kappa) f_l & \mathrm{for} \ l \neq 0 \end{array}
                \right.
\end{equation}
where $f_l = (\sqrt{3}-2)^{\modulus{l}}/\sqrt{3}$, $f_l = f_{-l}$ and $\kappa = \gamma_B / \gamma_A$ (see App.~\ref{app:DissipationWannier} for details).

Whilst the reservoirs of the A and B sites are independent, in the Wannier basis adjacent states share a common reservoir which enables a non-local purely dissipative coupling.  In  turn,  when $\kappa =  1$, the non-local dissipation terms $\gamma_l$ for $l \neq 0$ vanish and individual Wannier states decouple.

\section{Non-Interacting Regime}
\label{sec:NonInteracting}

In the absence of interactions, the master equation Eq.~(\ref{eq:site_master_equation}) can be solved using Ehrenfest's theorem $d \langle \hat{O} \rangle / dt = \langle \hat{O} \dot\rho \rangle = \mathrm{Tr} \{-i \hat{O} [H, \rho] + \hat{O} D[W] \rho\}$.  As the solutions of Eq.~(\ref{eq:site_master_equation}) are Gaussian states, they are entirely specified by the first and second order moments of the Wannier operators $W_i$.  Using the non-interacting version of the master equation, the equations of motion for the expectation values for $W_i$ and $W_i^\dag W_{i+j}$ are given by
\begin{align}
    \label{eq:Field_Expectation}
    \frac{d \langle W_i \rangle}{d t} & = - \frac{i}{2} \Omega^*_{W} \delta_{i,0} - \sum_l \gamma_l \expectation{W_{i+l}} \\
    \frac{d \langle W^\dag_i W_{i+j} \rangle}{d t} & = \frac{i}{2} \langle \Omega_{W} \delta_{i,0} W_{i+j} - \Omega^*_W \delta_{i+j,0} W^\dag_i \rangle \nonumber \\
    \label{eq:Wannier_Current}
    & - \frac{1}{2} \sum_l \gamma_l \left( \langle W^\dag_{i+l} W_{i+j} \rangle + \langle W^\dag_{i} W_{i+j-l} \rangle \right)
\end{align}
where $\delta_{i,j}$ is the Kronecker delta. Eqs.~(\ref{eq:Field_Expectation}) and~(\ref{eq:Wannier_Current}) provide a set of coupled equations that we solve numerically.  

The non-local dissipation coefficients $\gamma_l$ decrease exponentially with $l$ so we introduce a cutoff at $l > 10$ for which we set $\gamma_l = 0$.  Also, we expect long range correlations to be negligible so we make the approximation $\langle W^\dag_i W_{i+j} \rangle = 0$ for $j > 10$.  We have checked that the results presented here are converged with respect to these cutoffs.  A finite size lattice was used which was sufficiently large for boundary effects to be insignificant.

\begin{figure*}
  \subfigure{\includegraphics[width = 5.6cm]{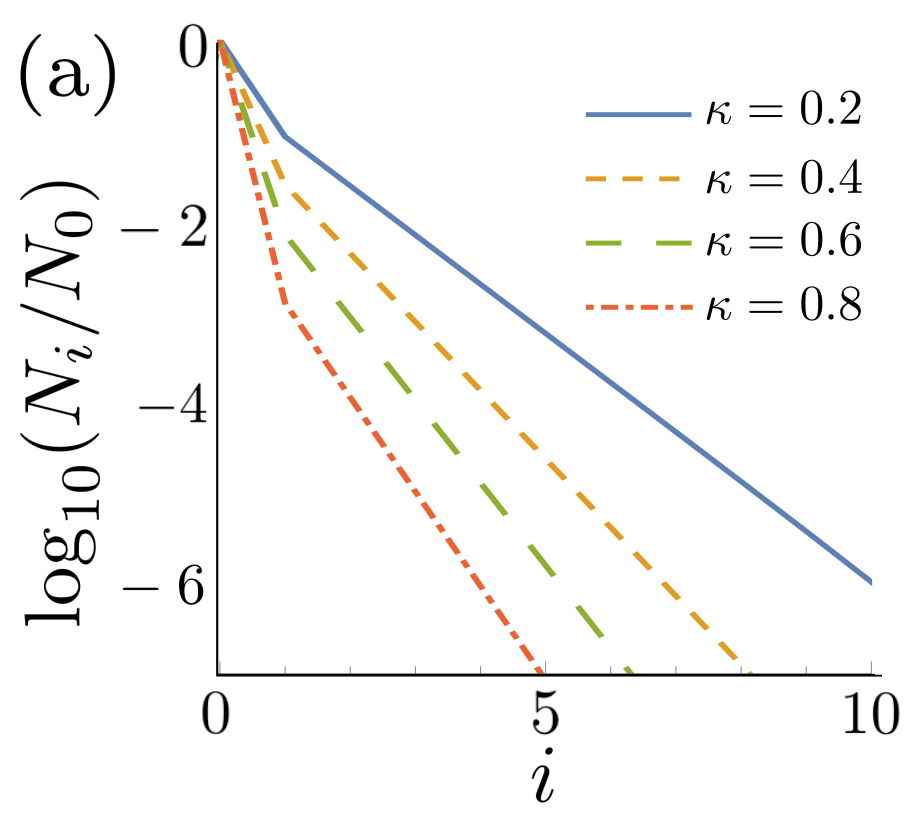}}
  \subfigure{\includegraphics[width = 5.6cm]{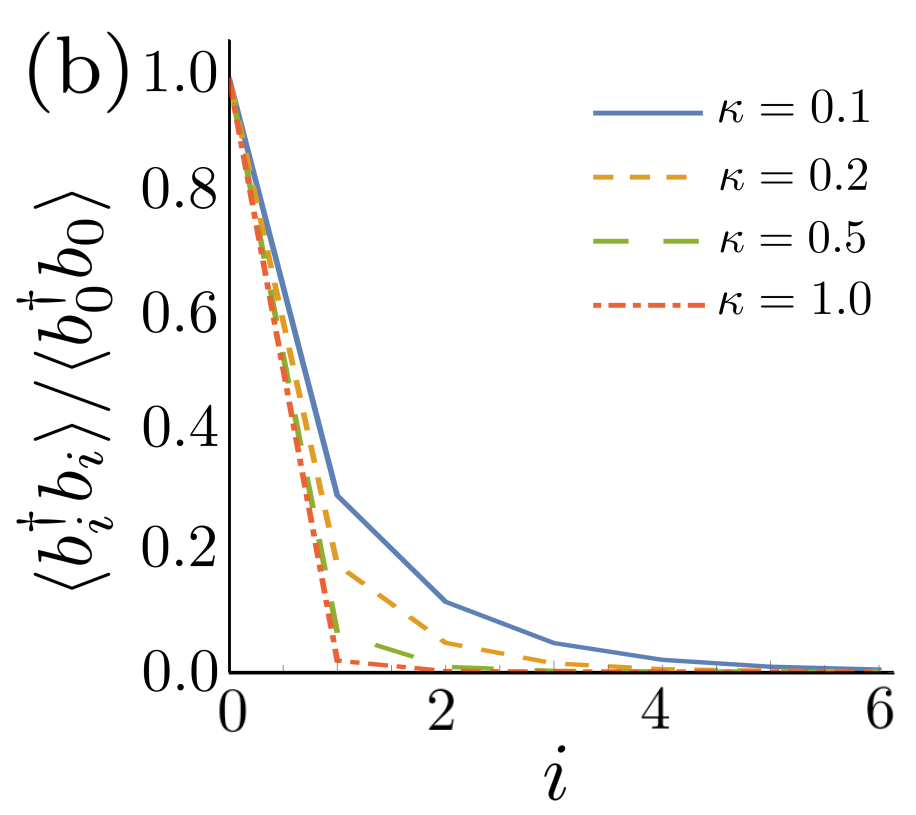}}
  \subfigure{\includegraphics[width = 5.6cm]{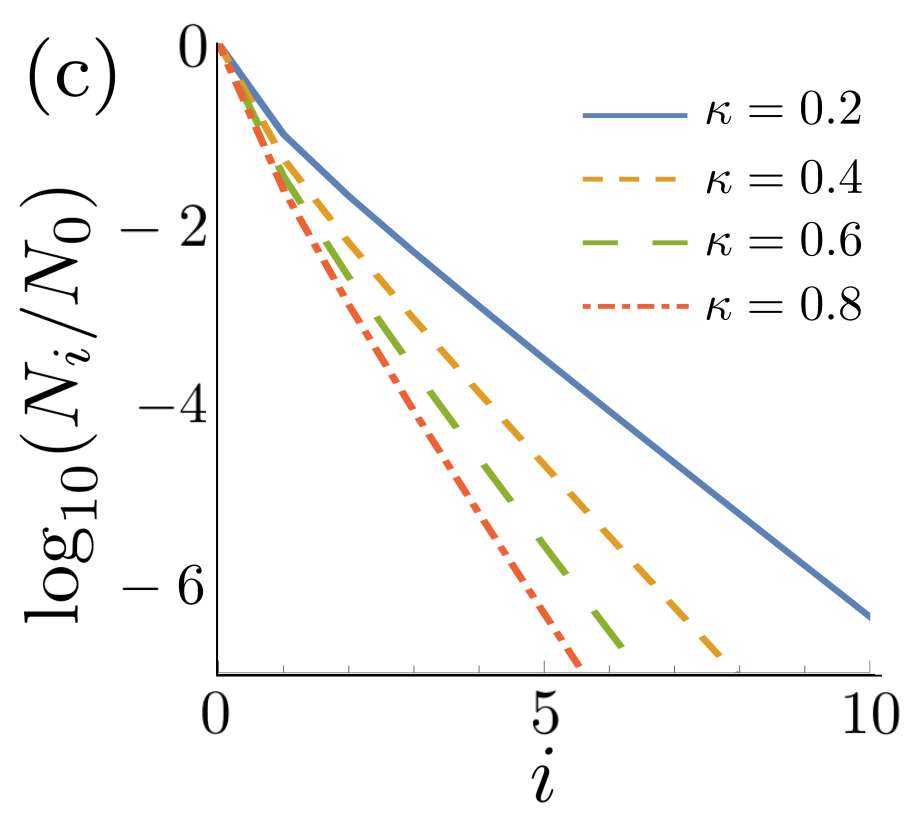}}
  \caption{(Color online) The non-interacting regime. (a) Normalised steady-state density of excitations in the Wannier basis for a coherent drive with $\Omega_W = \gamma_{A}$ at Wannier state $i = 0$. (b) Normalised steady-state density for the B sublattice. (c) Normalised steady-state density for an incoherent pump with $P_{B,0} = \gamma_{A}/100$ at $i = 0$.}
  \label{fig:full_panel}
\end{figure*}

In order to investigate dissipation-induced mobility in this frustrated lattice, we pump only the site at $i = 0$.  The advantage of pumping in the Wannier basis over the site basis is that it is easier to quantitatively disentangle the effect of the non-local dissipator.  Pumping a single site on the sawtooth lattice results in a pump profile of finite width in the Wannier basis which obscures the propagation of the excitations enabled by the dissipation.

\subsection{Dissipation-induced mobility}

Fig.~\ref{fig:full_panel}a shows the density of Wannier state excitations $N_i = \langle W^\dag_i W_i \rangle$ normalised by the density on the pumped site on a logarithmic scale.  Strikingly, we see that the excitations move away from the driven site even though there are no coherent processes which couple the Wannier states.  The transport in this system is enabled by the dissipator, which is achieved solely through coupling of the frustrated sawtooth lattice to local Markovian baths.  We will show that this mobility cannot be explained in terms of a diffusion process in Sec~\ref{sec:MobilityModelling}.

This effect can also be clearly seen in the excitation densities of the original lattice. Fig.~\ref{fig:full_panel}b shows the density on the B sites $\langle b_i^\dag b_i \rangle$ for the same drive intensity pattern as above.  In the lattice site basis, the drive amplitudes $\Omega_{A, i}$ and $\Omega_{B, i}$ are nonzero for $i \neq 0$ as we are pumping a single Wannier state.  However, while the excitation density does not vanish on the sites neighbouring the central site when there is no non-local dissipation ($\kappa = 1$), it is clear that the density profile changes significantly as the non-local dissipation terms increase.  This demonstrates that the mobility of the excitations is indeed due to dissipation-enabled transport and that the effect is not an artifact of the transformation into the Wannier basis.

\begin{figure}
  \includegraphics[width=\columnwidth]{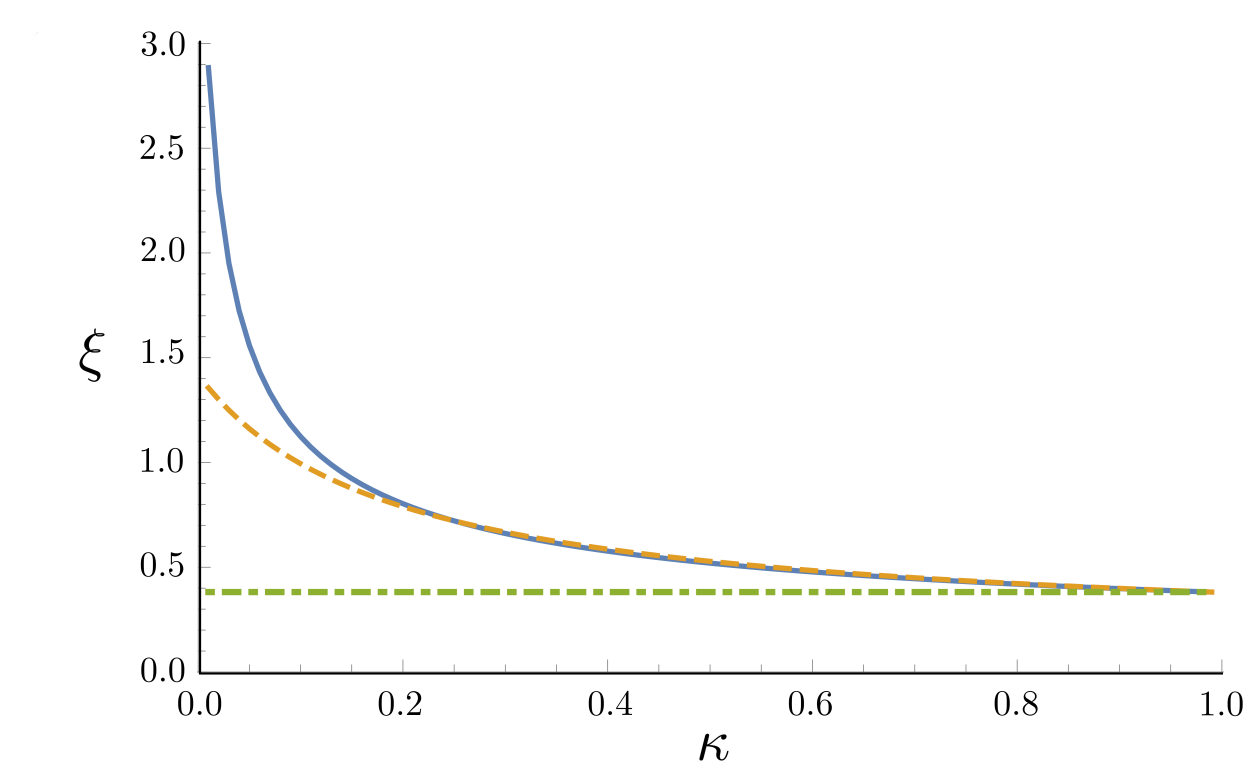}
  \caption{(Color online) Decay length of the density profile $\xi$ as a function of $\kappa$ for direct coupling (dot-dashed), effective drive (dashed) and exact (solid line) calculations where the Wannier state $W_0$ is driven coherently with $\Omega_W = \gamma_{A}$.}
  \label{fig:decay_length}
\end{figure}

The normalised density decreases exponentially and can be approximated by $ N_i = \exp (- |r_i| / \xi)$ for $i \neq 0$.  The decay length of the density profile can thus be extracted using adjacent densities $\xi_i = \left|\log_{10} N_i - \log_{10} N_{i+1}\right|^{-1}$ and is shown in Fig.~\ref{fig:decay_length} as a function of $\kappa$ for $i = 4$.  As $\kappa$ decreases, the non-local dissipation rates $\gamma_l$ increase which leads to a divergence in $\xi$ as $\kappa \to 0$. In this limit, the dissipation rates on the B sublattice tend to zero which leads to the formation of the dark state $\sum_i (-1)^i W_i^\dag \ket{0} = \sum_i (-1)^i b_i^\dag \ket{0}$. The decay length of the density of excitations in the lattice introduced by the pump diverges as this dark state extends over the entire lattice.  In the limit $\kappa \to \infty$, where the dissipation rate on the A sites tends to zero, the corresponding dark state $\sum_i (-1)^i a_i^\dag \ket{0}$ would involve contributions from the dispersive band and is thus not accessible with our driving mechanism.

\subsection{Long-range first-order coherences}

The local dissipation in the site basis can also generate first order correlations $\langle W^\dag_j W_l \rangle$ in the sawtooth lattice. Remarkably such coherences even emerge when the coherent input is removed $\Omega_W = 0$ and the system is incoherently pumped. To describe the incoherent pumping, we add a term $\sum_i \sum_{x=a,b} (P_{x, i}/2) \left(2 x_i^\dag \rho x_i - \{x_i x_i^\dag, \rho \}\right)$ to the master equation (\ref{eq:site_master_equation}) where $P_{x, i}$ is the strength of the incoherent pump.  The density profile generated by a local incoherent pump is shown in Fig.~\ref{fig:full_panel}c and exhibits similar exponentially decaying extension as for a coherent input. Numerical simulations show that the first order coherence $g^{(1)}(j,l) = \langle W^\dag_j W_l \rangle / \sqrt{\langle W^\dag_j W_j \rangle \langle W^\dag_l W_l \rangle}$ for coherent driving of the Wannier state $W_0$ is perfectly correlated with $g^{(1)} (j, l) = (-1)^{|j-l|}$ as the density matrix of the system is a product of coherent states with alternating phases.  We thus find long-range first-order coherence despite an exponentially decaying spatial density profile. This indicates that the excitations form a condensate whose extension is the decay length $\xi$ of the density profile. 

\begin{figure}
  \includegraphics[width=\columnwidth]{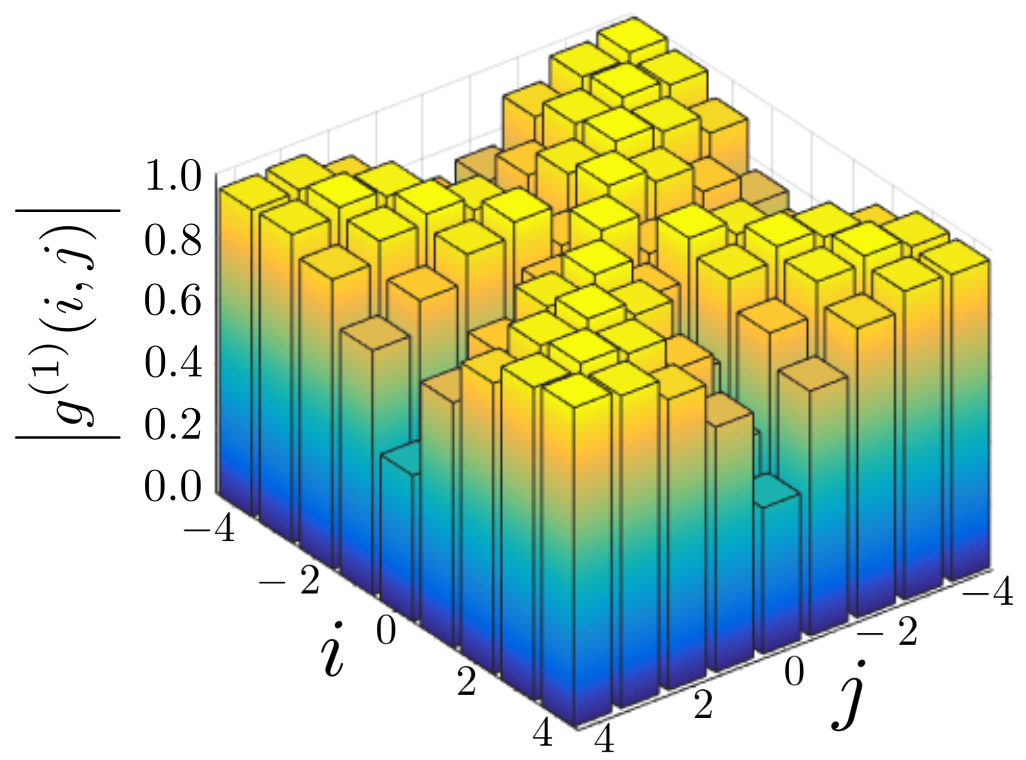}
  \caption{(Color online) Spatial coherences in the steady state $g^{(1)}(j, l)$ for an incoherent pump at $i = 0$ with $P_{B, 0} = \gamma_A / 100$ and $\kappa = 0.1$.}
  \label{fig:g1_incoherent}
\end{figure}

The same function for a system where the coherent input is replaced with an incoherent drive on the B sublattice at site $i = 0$ is shown in  Fig.~\ref{fig:g1_incoherent}.  The dissipation-induced mobility continues to generate significant coherences even in the absence of a coherent input.  As demonstrated for the coherent case, mobile excitations propagate coherently through the lattice.  However, an incoherent drive also counteracts the build-up of the first-order coherence $g^{(1)} (j, l)$.  As the incoherent pump is strongest for the Wannier site $W_0$, the coherence is reduced around $g^{(1)} (0, l)$ as seen in Fig.~\ref{fig:g1_incoherent}.  In contrast to recent experimental observations~\cite{Baboux:2016}, the coherences in our system are thus longer range, due to the absence of disorder. 

\section{Modelling the decay length of the dissipation-induced mobility}
\label{sec:MobilityModelling}

\begin{figure}
    \centering
    \includegraphics[width=\columnwidth]{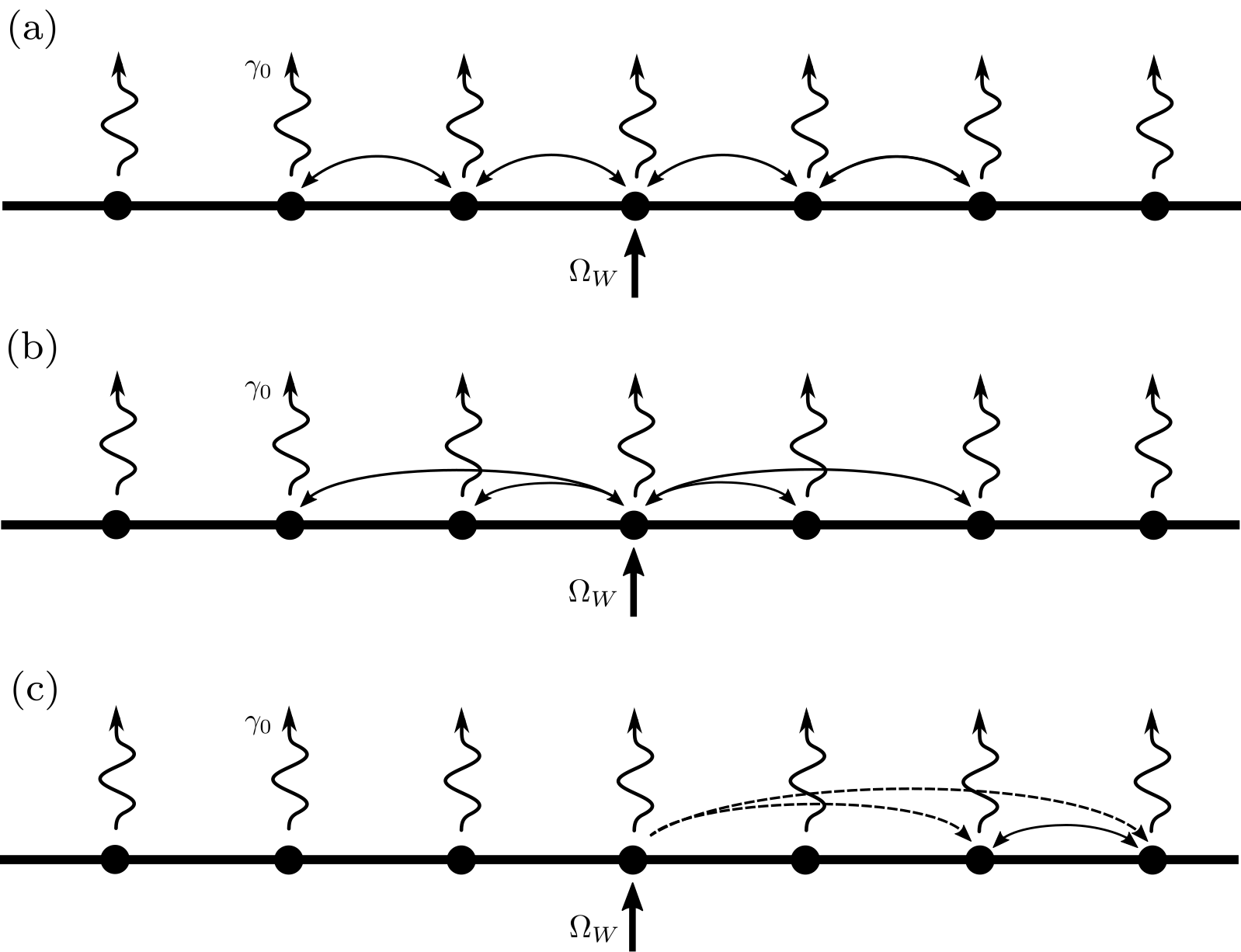}
    \caption{Diagrammatic representations of the models used to explain dissipation-induced mobility in the non-interacting regime.  The central site is pumped with a drive of strength $\Omega_{W, 0}$ and there is a dissipation rate $\gamma_0$ for each site.  (a) The single site diffusion model assumes that only tunnelling between neighbouring sites is important.  (b) The direct non-local dissipative coupling model, where the density on each site is modelled by a two-site model consisting of the relevant site and the pumped site coupled by the non-local dissipation term.  (c) The effective drive model consists of a three site model where the pumped site is not affected by the two remote, coupled, unpumped sites.}
    \label{fig:model_diagrams}
\end{figure}

In this section we clarify that the dissipation-induced mobility which we find cannot be attributed to a diffusion process, see Fig.~\ref{fig:model_diagrams}a, as this fails to describe the decay length of the density correctly.  We continue by examining which of the dissipative transfer processes of Eq.~(\ref{eq:site_master_equation}) contribute significantly to the mobility by comparing the predicted density distributions for two approximate models to our numerical results.

Considering only the direct transfer of excitations from the pumped site to unpumped remote sites through long-range non-local dissipative incoherent tunnelling events, see Fig.~\ref{fig:model_diagrams}b, is not sufficient to explain our findings as it predicts that the decay length $\xi$ is independent of $\kappa$.  A refined version of this model, whereby a corrected density on the pumped site acts as an effective drive for two remote sites with nearest-neighbour hopping, see Fig.~\ref{fig:model_diagrams}c, provides a good approximation for the density decay length $\xi$ for a wide range of $\kappa$. The model breaks down as $\kappa \to 0$ where the contribution of the longer range non-local dissipation processes which we have ignored becomes significant.

\subsection{Single site diffusion model}

We begin by attempting to model the density profile by approximating the motion as a diffusive random walk where excitations can hop incoherently between neighbouring sites, see Fig.~\ref{fig:model_diagrams}a.  The continuity equation for the density on site $i$ for a driven-dissipative system is given by
\begin{equation}
    \label{eq:continuity}
    \frac{\partial N_i}{\partial t} = - \left. \frac{\partial N_i}{\partial t} \right|_L - \left. \frac{\partial N_i}{\partial t} \right|_R - \Gamma N_i + F (x_i)
\end{equation}
where $\partial N_i / \partial t |_{L/R}$ is the number of particles on site $i$ moving to the left and right respectively, $\Gamma$ is the dissipation rate and $F (x_i)$ is a source distribution.  The net particle flow to the site $i+1$ in a time interval $\Delta \tau$ is given by
\begin{align}
    \left. \frac{\partial N_i}{\partial t} \right|_R & = \frac{J (N_i - N_{i+1})}{\Delta \tau} \\
                                    & = - J \frac{\Delta x}{\Delta \tau} \frac{\partial N_i}{\partial x}
\end{align}
where $J$ is the rate at which particles hop between adjacent sites and similarly for $\partial N_i / \partial t |_L$.  Therefore, the diffusion equation for Eq.~(\ref{eq:continuity}) is given by
\begin{equation}
    \label{eq:diffusion}
    \frac{\partial N_i}{\partial t} = J \frac{\Delta x^2}{\Delta t} \frac{\partial^2 N_i}{\partial x^2} - \Gamma N_i + F (x_i)
\end{equation}
where we note that $\Delta x^2 / \Delta \tau$ is some scaled diffusion constant $D$.  The steady-state solution of Eq.~(\ref{eq:diffusion}) without sources is given by $N_i = A e^{- x_i / \Xi} + B e^{ x_i / \Xi}$ with the decay length $\Xi = \sqrt{J D / \Gamma}$.  

Assuming that the incoherent dissipative coupling can be modelled as a diffusive process, the most natural values for the dissipation and hopping rates are $\Gamma \propto \gamma_0$ and $J \propto \gamma_1$.  Therefore, the functional form of the $\Xi$ would be
\begin{equation}
    \Xi (\kappa) \propto \sqrt{\frac{\gamma_1}{\gamma_0}} \propto \left(\frac{2 f_0 + f_1}{f_0 (1 - \kappa)} - 1 \right)^{- \frac{1}{2}}
\end{equation}
which differs from the $\xi$ obtained from the numerical simulations.  If we include the direct non-local dissipative coupling of the pumped site to remote unpumped sites as a source term for Eq.~(\ref{eq:continuity}), which will have a functional form $F (x_i) \propto e^{-\modulus{x_i}/\xi_\Omega}$ where $\xi_\Omega$ is the decay length of the direct coupling, then this will affect the steady-state density by introducing a particular integral with decay length $\xi_\Omega$ which will not affect the functional form of $\Xi (\kappa)$.  Indeed, this will prevent the density distribution from having the exponential distribution observed in the simulations.

\subsection{Direct non-local dissipative coupling model}

As the dissipation-induced mobility we observe cannot be understood in terms of diffusion through nearest-neighbour hopping we will now examine which dissipative transfer processes play an important role.  The non-local dissipation is long ranged so a simple approximation is to neglect the dissipative coupling between sites which are not pumped such that transport from the pumped site $i = 0$ to another site $i = j$ is solely mediated by direct non-local dissipative coupling $\gamma_j$ between the central pumped site and the remote unpumped sites, see Fig.~\ref{fig:model_diagrams}b.  

We can calculate the density distribution for this model by solving Eqs.~(\ref{eq:Field_Expectation}) and~(\ref{eq:Wannier_Current}) for a hypothetical two-site system consisting of the sites $i = 0$ and $i = j$ only.  The zeroth site is pumped with a drive strength $\Omega_W$ so these equations give the steady-state density on the $j$th site
\begin{equation}
    \label{eq:direct_twostate_dens}
    \expectation{W_j^\dag W_j} = \frac{\gamma_j^2 \modulus{\Omega_W}^2}{4 (\gamma_0^2 - \gamma_j^2)^2} \approx \frac{\gamma_j^2 \modulus{\Omega_W}^2}{4 \gamma_0^4}
\end{equation}
where $\gamma_0 \gg \gamma_j$ due to the exponential localisation of the Wannier states so the decay length is given by
\begin{equation}
    \xi_\Omega \approx \left[ \log_{10} \left(\frac{\gamma_{j+1}^2}{\gamma_j^2}\right) \right]^{-1} \approx 0.38
\end{equation}
In the limit $\kappa \to 1$, the non-local dissipation rates vanish and the density distribution is predicted well by this model.

\subsection{Effective drive model}
\label{subsec:EffectiveDrive}

The direct coupling model was accurate in the limit $\kappa \to 1$ so we can hope that a more accurate model will be obtained by incorporating nearest neighbour hopping.  Therefore, we solve the equations of motion for the sites $i = \{0, j, j+1\}$ where $j > 1$ assuming that these two sites are not directly coupled to the pumping site with nearest neighbour dissipative coupling terms, see Fig.~\ref{fig:model_diagrams}c.  We make the assumption that for $i = \{j, j+1\}$ the only relevant term in Eq.~(\ref{eq:Field_Expectation}) is from the pumped site $l = - j$ such that $\sum_{l \neq \{0, 1\} } \gamma_l \expectation{W_{j + l}} \approx \gamma_{-j} \expectation{W_0}$.  Furthermore, we assume that the field on the pumped site is only affected by its nearest neighbours so we can treat $\expectation{W_0}$ as a constant.

We thus calculate the amplitude of the pumped site $\expectation{W_0}$ by solving Eq.~(\ref{eq:Field_Expectation}) for the pumped site and its nearest neighbours and use this result to calculate the densities on sites $i = j$ and $i = j+1$.  Details can be found in App.~\ref{app:DecayRateDerivation} and result in a decay length of 
\begin{equation}
    \xi \approx \left[2 \log_{10} \left(\frac{\gamma_0 \gamma_j - \gamma_1 \gamma_{j+1}}{\gamma_1 \gamma_j - \gamma_0 \gamma_{j+1}} \right)\right]^{-1}
\end{equation}
As we see in Fig.~\ref{fig:decay_length}, this approximation works well for a wide range of non-local dissipation strengths.  However, as $\kappa \to 0$, the approximation breaks down and the decay length is underestimated due to multiple hopping processes which have not been included.

We thus conclude that for $\kappa \approx 1$, the direct process from the pumped site to the considered site plays the most significant role.  As $\kappa$ decreases, the transfer of particles from the neighbouring site to the considered site becomes relevant.  The smaller $\kappa$ becomes, the more these indirect processes play a role and the more sites must be included to provide an accurate approximation of the density distribution.

\section{Strongly-Interacting Regime}  
\label{sec:StronglyInteracting}

After investigating the non-interacting case, we now turn to explore the effect of interactions on dissipation-induced mobility and coherence.  As with the dissipator, the projection of the interaction Hamiltonian $H_U$ onto the Wannier basis results in non-local interaction terms, see App.~\ref{app:IntCoeffCalc} for a full derivation. These non-local interactions can also cause propagation, as has been explored in detail~\cite{Huber:2010, Takayoshi:2013, Tovmasyan:2013, Phillips:2015, Pudleiner:2015}.

The interaction induced terms which cause coherent propagation are least effective in the limit of very strong interaction, where propagation from terms such as $(W_{i}^{\dag})^{2}(W_{i+1})^{2}$ is suppressed.  To explore the effects of the dissipation, we thus consider this limit, where the dominant on-site interactions suppress multiple Wannier state excitations, and for simplicity set $U_A = 0$.  Truncating the system to the single-excitation subspace, the Hamiltonian in Eq.~(\ref{eq:Interaction}) in the Wannier basis can be approximated by
\begin{align}
    H_U \approx & \sum_i [ U_0 W^\dag_i W^\dag_i W_i W_i + U_1 W^\dag_i W_i W^\dag_{i+1} W_{i+1} \nonumber \\
    & \qquad + U_2 W^\dag_i W_i W^\dag_{i+2} W_{i+2} \nonumber \\
    & \qquad + U_3 (W^\dag_{i-1} W^\dag_i W_i W_{i+1} + \mathrm{h. c.})]   
\end{align}
where $U_0 \approx 0.192 U_B$, $U_1 \approx 0.133 U_B$, $U_2 \approx 0.054 U_B$ and $U_3 \approx 0.023 U_B$ are effective interaction strengths derived in App.~\ref{app:IntCoeffCalc}.  The leading-order contributions of $H_U$ are the onsite interaction, nearest-neighbour cross-Kerr interactions, which cannot move an excitation from one site to another, and density-assisted tunnelling, whereby an excitation can coherently tunnel if there is an excitation on a nearby site.  The density-assisted tunnelling terms contribute to the mobility for our driven-dissipative sawtooth lattice and could obscure or even dominate over the contribution induced by the non-local dissipator.  We now show that the joint effect of the interaction terms suppresses the mobility in the strong interaction regime and that, even here, the main contribution to the mobility originates from the dissipation.  To this end, we calculated the density distribution in the strongly interacting regime using a variational matrix product operator (MPO) method \cite{Schollwock:2011, Mascarenhas:2015, Cui:2015} (see App.~\ref{app:MPODetails}).

\begin{figure}
    \includegraphics[width=\columnwidth]{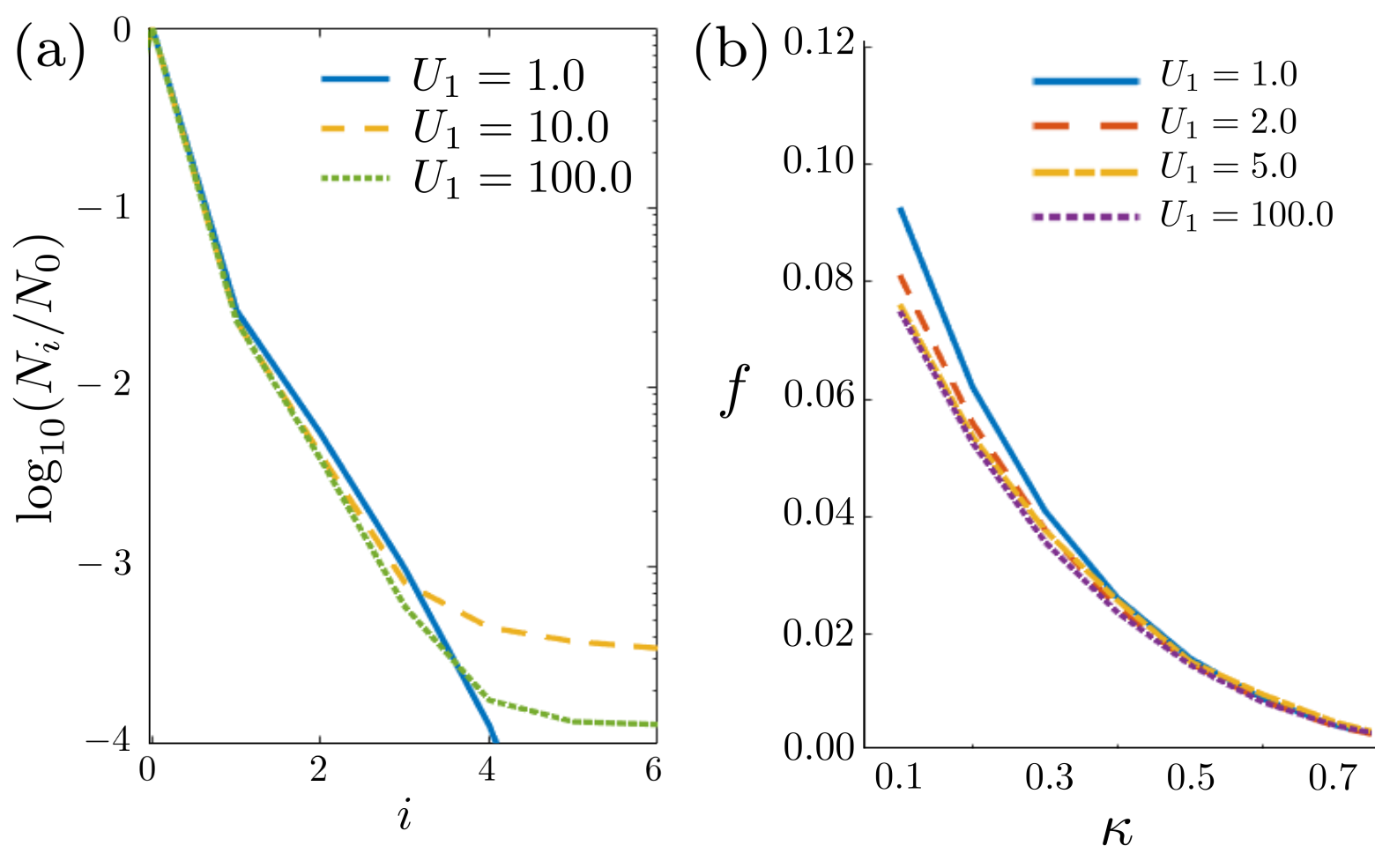}
    \caption{(Color online) (a) Normalised steady-state density of excitations in the Wannier basis for the interacting regime and a single driven Wannier state at $i = 0$, $\Omega_{W, 0} = \gamma_{A}$ and $\kappa = 0.2$ (c.f. Fig.~\ref{fig:full_panel}a). (b) Fraction of excitations not on the pumped site $f$ as a function of the dissipation rate asymmetry ratio $\kappa$ for an interacting system with a cross-Kerr interaction strength $U_1$. }
    \label{fig:density_profile_withinteractions}
\end{figure}

\begin{figure}
  \includegraphics[width=\columnwidth]{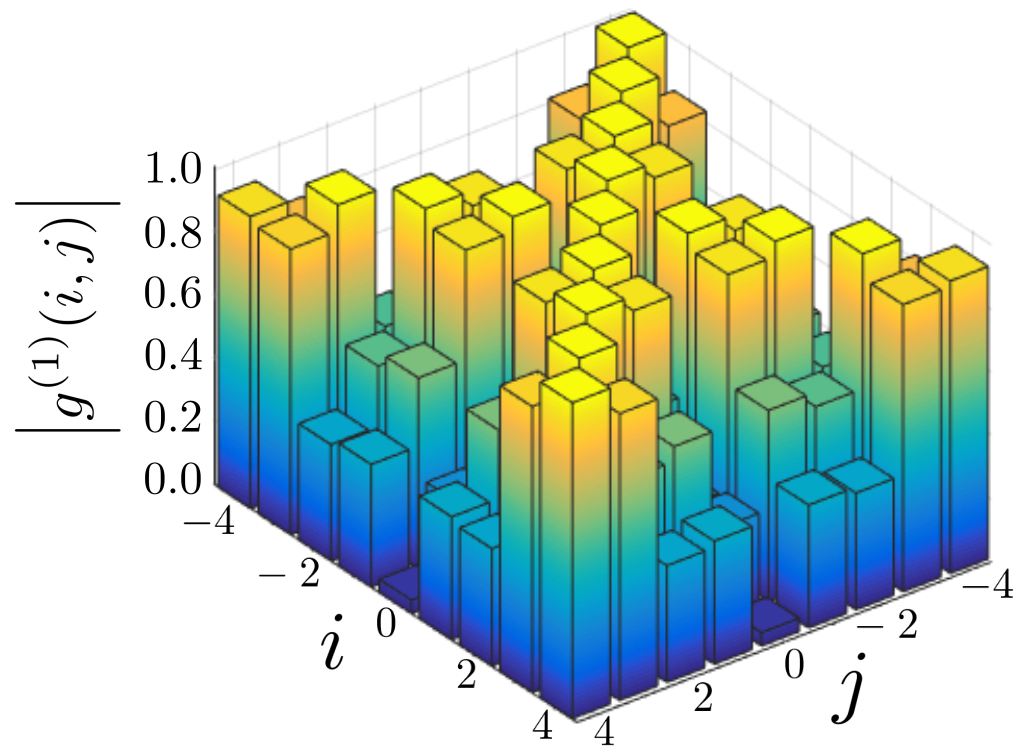}
  \caption{(Color online) Spatial coherences in the steady state $g^{(1)}(j, l)$ for a coherent drive at $i = 0$ with $\Omega_W = \gamma_A$ and $\kappa = 0.1$ in the strongly-interacting regime with $U_1 = 100$.}
  \label{fig:g1_quantum}
\end{figure}

The interactions modify the density distribution which, in contrast to the non-interacting case, no longer decays exponentially away from the pumped site, see Fig.~\ref{fig:density_profile_withinteractions}a.  To study how $H_U$ affects the dissipation-induced mobility, we thus use the fraction of excitations in the unpumped Wannier states $f = (\sum_{i \neq 0} N_i) / (\sum_i N_i)$ as the relevant figure of merit ($f = 0$ for uncoupled Wannier states).  Fig.~\ref{fig:density_profile_withinteractions}b shows this non-local density fraction as a function of $\kappa$ for a range of $U_1$ for which the truncation to the single-excitation subspace is valid (see App.~\ref{app:SingleExcitationSubspace}).  For fixed $\kappa$, the mobility of the excitations decreases with increasing interaction strength. We attribute this reduction in the mobility to the cross-Kerr interactions, which shift the energy of the neighbouring site, detuning the non-local dissipative transition and preventing excitations from moving between sites.  The effect of the density-assisted tunnelling terms is insignificant as setting $U_2 = 0$ leads to a variation of $f$ of the order of $10^{-2}$. Moreover, the dependence of $f$ on $U_1$ is much weaker than its dependence on $\kappa$, showing that the non-local dissipator is the dominant contributor to the steady-state of the system in the limit of strong interactions.  We also computed the coherences between lattice sites for $U_1 = 100$, $\Omega_W = \gamma_A$ and $\kappa = 0.1$, see Fig.~\ref{fig:g1_quantum}.  Similarly to the incoherent drive, interactions counteract the build-up of first order coherence. Since the density of excitations is highest for the pumped site, interactions are most effective on this site, which reduces $g^{(1)} (j, l)$ around $j = 0$.  Our results demonstrate that the generation of coherences due to dissipation-induced mobility is a general phenomenon and not restricted to non-interacting systems.

\section{Conclusions}
\label{sec:Conclusions}

We have shown that local Markovian dissipation can induce mobility and long range coherence in frustrated lattice systems in the absence of kinetic energy.   An experimental demonstration of the effect we predict here could be realised in any driven-dissipative lattice with a flat band. For example, in photonic waveguides lattices~\cite{Mukherjee:2015A}, the couplings between sites can be engineered to realise a flat band system where defects and disorder can be kept low enough to not affect the dynamics of the lattice excitations.  Neighbouring waveguides can be used to simulate the local Markovian baths with a strong degree of control over the environmental dynamics as has recently been demonstrated~\cite{Mukherjee:2017}.  Additionally, microwave resonator arrays in superconducting platforms with their exquisite precision in fabrication and control provide a suitable system.  An imbalance in the dissipation rates of A and B sublattices is expected, or can be engineered, in many situations and could be made stronger via Purcell enhancements.  Whilst we have discussed steady-state density distributions, the effects we find could also be studied dynamically, where varying the imbalance in dissipation rates would change the rate at which a localised excitation disperses across the lattice.  Photonic waveguide lattices~\cite{Mukherjee:2015A} would be particularly well suited for such experiments.

\section{Acknowledgements}

We thank Patrik \"{O}hberg, Erik Gauger and Manuel Valiente for useful discussions. This work was supported by EPSRC under grant No. EP/N009428/1. O.T.B. acknowledges support from EPSRC CM-CDT Grant No. EP/G03673X/1.

\appendix

\section{Dissipation in the Wannier basis for the Sawtooth Lattice}
\label{app:DissipationWannier}

The equations of motion for the driven-dissipative sawtooth lattice is given by the master equation $\dot{\rho} = - i [H, \rho] + \frac{1}{2} \sum_i \left( \gamma_{A, i} D[a_i] \rho + \gamma_{B, i} D[b_i] \rho \right) $ where the dissipator $D[a_i] \rho = 2 a_i \rho a^\dag_i - \{a^\dag_i a_i, \rho\}$ describes Markovian loss of an excitation from site $i$.  For simplicity, we will assume that the losses from all the A sites and all B sites are the same but the loss rates for A and B sites are not necessarily equal ie. $\gamma_{A, i} = \gamma_A$ and $\gamma_{B, i} = \gamma_B$.

If we take Eq.~(\ref{eq:a_operator_def}) and plug it into the dissipator $\sum_i D[a_i] \rho$, we get
\begin{align}
    \sum_i D[a_i] \rho = & \sum_i \gamma_A (2 a_i \rho a^\dag_i - \{a^\dag_i a_i, \rho\}) \\
                        = & \sum_{i,j,k} \gamma_A w_A(r_i - r_j) w_A (r_i - r_k) \nonumber \\
                        & \qquad \cdot(2 W_j \rho W^\dag_k - \{W^\dag_j W_k, \rho \}) 
\end{align}
where we note that $w_A(r)$ is real as it is a Fourier transform of a symmetric function.  Similarly, for the dissipator $\sum_i D[b_i] \rho$,
\begin{align}
    \sum_i D[b_i] \rho = & \sum_i \gamma_B (2 b_i \rho b^\dag_i - \{b^\dag_i b_i, \rho\}) \\
                        = & \sum_{i,j,k} \gamma_B w_B(r_i - r_j) w_B (r_i - r_k) \nonumber \\
                        & \qquad \cdot(2 W_j \rho W^\dag_k - \{W^\dag_k W_j, \rho \}) 
\end{align}
Combining these two equations, we have
\begin{multline}
    \sum_i (\gamma_A D[a_i] + \gamma_B D[b_i]) \rho = \\
    =\sum_{j, l} \gamma_{j,l}^{\mathrm{eff}} (2 W_j \rho W^\dag_{j+l} - \{W^\dag_{j+l} W_j, \rho \}) 
\end{multline}
where $\gamma_{j,l}^{\mathrm{eff}} = \gamma_l^{\mathrm{eff}}$ only depends on the separation $l$ of the Wannier states and
\begin{align}
    \gamma_l^{\mathrm{eff}} = & \sum_{i'} \gamma_A  w_A(r_{i'}) w_A (r_{i'} - r_{l})\nonumber \\
    & \qquad \qquad + \gamma_B w_B(r_{i'}) w_B (r_{i'} - r_{l})) \\
    = & \frac{1}{2 \pi} \int_{-\pi}^\pi \frac{(2 \gamma_A \cos^2 (k/2) + \gamma_B) }{\cos k + 2} e^{i k l} \dd k \\
    = & \frac{1}{2 \pi} \int_{-\pi}^\pi \frac{\gamma_A (1 + \cos k) + \gamma_B}{\cos k + 2} \cos (k l) \dd k \\
    = & \frac{1}{2 \pi} \int_{-\pi}^\pi (\gamma_A + \gamma_B) \frac{\cos (k l) }{\cos k + 2} \dd k \nonumber \\
      & \qquad + \frac{1}{2 \pi} \int_{-\pi}^\pi \frac{\gamma_A}{2} \frac{\cos k(l+1) + \cos k(l-1)}{\cos k + 2} \dd k \\
    = & \frac{1}{2} \left( \gamma_A (2 f_l + f_{l-1} + f_{l+1}) + 2 \gamma_B f_l \right)
\end{align}
If $\gamma_A$ and $\gamma_B$ are equal, we can take the sum over $i$ which gives a local dissipator due to the Wannier state orthogonality relation
\begin{equation}
    \label{eq:Wannier_orthogonality}
    \sum_i w_A (r_j - r_i) w_A (r_k - r_i) + w_B (r_j - r_i) w_B (r_k - r_i) = \delta_{jk} 
\end{equation}
and all non-local dissipation terms disappear.  When the dissipation rates are not equal, we can evaluate the integral $f_l$ using the residue theorem.  For simplicity, let us assume that $l \geq 0$ and we note that $f_l = f_{-l}$.  First, we make the standard substitution $z = e^{i k}$ such that
\begin{equation}
    \label{eq:residueintegral}
    \int_{-\pi}^\pi \frac{\cos k l}{\cos k + 2} \dd k = \int_{\mathcal{C}} \frac{z^{2l} + 1}{z^2 + 4z + 1} \frac{\dd z}{i z^l}
\end{equation}
where the integration contour $\mathcal{C}$ runs anticlockwise around the unit circle.  This integral has three poles: two simple poles at $z = - 2 \pm \sqrt{3}$ and a pole of order $l$ at $z = 0$.  The pole at $z = - \sqrt{3} - 2$ lies outside the unit circle so it does not contribute to the integral.  The residue at $z = \sqrt{3} - 2$ is calculated using the standard method for a simple pole
\begin{align}
    \mathop{\mathrm{Res}}_{z = \sqrt{3} - 2}  \frac{z^{2l} + 1}{z^l (z^2 + 4z + 1)} 
                & = \lim_{z \to \sqrt{3} - 2} \frac{z^{2l} + 1}{z^l (z + 2 + \sqrt{3})} \\
                & = \frac{(\sqrt{3} - 2)^l + (\sqrt{3} - 2)^{-l}}{2 \sqrt{3}}
\end{align}
For the pole at $z = 0$, we use the Laurent expansion of the integrand and take the prefactor of the $z^{-1}$ term.  The lowest order term in the Laurent expansion when $z^{2l}$ is in the numerator is of order $z^{l}$ so we only need
\begin{widetext}
\begin{align}
    \frac{1}{z^l} \frac{1}{z^2 + 4 z + 1} & = \frac{1}{z^l} \left( \frac{1}{1 - \frac{z}{- \sqrt{3} - 2}} \right) \left( \frac{1}{1 - \frac{z}{\sqrt{3} - 2}} \right) \\
    & = \frac{1}{z^l} \sum_{n=0}^\infty \left( \frac{z}{\sqrt{3} - 2} \right)^n \sum_{m=0}^\infty \left( \frac{z}{- \sqrt{3} - 2} \right)^m \\
    & = \frac{1}{z^l} \left( \ldots + \sum_{n=0}^{l-1} \frac{z^{l-1}}{(-\sqrt{3}-2)^n (\sqrt{3}-2)^{l-1-n}} + \ldots \right) \\
    & = \frac{1}{z^l} \left( \ldots + \frac{(\sqrt{3}-2)^{l} - (\sqrt{3}-2)^{-l}}{2 \sqrt{3}} z^{l-1} + \ldots \right) 
\end{align}
Therefore, the integral in Eq.~(\ref{eq:residueintegral}) can be evaluated using the residue theorem to give
\begin{align}
    \int_{-\pi}^\pi \frac{\cos k l}{\cos k + 2} \dd k & = 2 \pi \left(\frac{(\sqrt{3} - 2)^l + (\sqrt{3} - 2)^{-l}}{2 \sqrt{3}} + \frac{(\sqrt{3} - 2)^l - (\sqrt{3} - 2)^{-l}}{2 \sqrt{3}} \right) \\
    & = 2 \pi \frac{(\sqrt{3} - 2)^l}{\sqrt{3}}
\end{align}
\end{widetext}
so $f_l = (\sqrt{3}-2)^{\modulus{l}} / \sqrt{3}$ where the absolute value of $l$ is taken as we know that the integral is symmetric under the transformation $l \to -l$. 

For reference, some numerically integrated values of $f_l$ are given below:
\begin{equation}
    \begin{array}{l|c|c|c}
        l & f_l & \gamma^{\mathrm{eff}}_l (\kappa = 0.1) / \gamma_A & \gamma^{\mathrm{eff}}_l (\kappa = 0.5) / \gamma_A  \\
        \hline
        0 & 0.57735     & 0.4803840     & 0.7113240     \\
        1 & -0.154701   & 0.1392809     & 0.0773505     \\
        2 & 0.0414519   & -0.0373060    & -0.0207259    \\
        3 & -0.011107   & 0.0099963     & 0.0055535     \\
        4 & 0.00297611  & -0.0026785    & -0.0014881    \\
        5 & -0.000797447& 0.0007177     & 0.0003987 
    \end{array} \nonumber
\end{equation}
We see that the dissipator contains non-local terms but that the dissipation rate for coupling to adjacent sites drops off rapidly.  There are negative dissipation rates but the dissipation rate matrix $\gamma^\mathrm{eff}_{j, l}$ is positive.

To connect to the non-local dissipation coefficients for the Lieb lattice calculated in the following section of this appendix, we also note that $f_l$ can be expressed in terms of the regularised hypergeometric function $f_l = {}_3 \tilde{F}_2 (1/2, 1,1 ;1+l, 1-l; -2)$

\section{Wannier states of the Lieb Lattice}

In the main text, we have focused on the sawtooth lattice, which is the simplest one-dimensional frustrated system.  However, the conclusions we derive in this paper are equally applicable to other frustrated lattices where the master equation for the Wannier states of the flat band have different dissipation rates for the different sublattices, as in Eq.~(\ref{eq:site_master_equation}).  Notably, the Lieb lattice, studied in some contemporary works on driven-dissipative frustrated systems~\cite{Baboux:2016, Biondi:2015, Casteels:2016}, exhibits the same behaviour.

The Lieb lattice consists of a one-dimensional chain of harmonic oscillators with nearest-neighbour coupling $J$ decorated with a coupling $g$ to an additional harmonic oscillator for every other site in the chain.  The Hamiltonian for this system is given by
\begin{align}
    H = & \sum_i \omega_A a^\dag_i a_i + \omega_B b_i^\dag b_i + \omega_C c_i^\dag c_i \nonumber \\
        & \quad + J \left(a^\dag_{i-1} b_i + a^\dag_i b_i + \mathrm{h.c.} \right) \nonumber \\
        \label{eq:HLieb_sitebasis}
        & \quad + g \left(a^\dag_i c_i + \mathrm{h.c.} \right)
\end{align}
where the unit cell is labelled with A, B and C sites, $\omega_A$, $\omega_B$ and $\omega_C$ are their respective onsite energies and $a_i$, $b_i$ and $c_i$ are the annihilation operators for the respective sublattices of the unit cell $i$.  Transforming Eq.~(\ref{eq:HLieb_sitebasis}) into the momentum basis and expressing distances in units of the lattice spacing one gets
\begin{equation}
    \label{eq:HLieb_momentumbasis}
    H = \sum_k [a^\dag_k \ b^\dag_k \ c^\dag_k] \left[ \begin{array}{ccc}
         \omega_A & J (1 + e^{-i k}) & g  \\
         J (1+e^{i k}) & \omega_B & 0 \\
         g & 0 & \omega_C \end{array}
         \right] \left[ \begin{array}{c} a_k \\ b_k \\ c_k \end{array} \right]
\end{equation}
When $\omega_B = \omega_C$, the dispersion relation of Eq.~(\ref{eq:HLieb_momentumbasis}) contains a flat band at $E = \omega_B$ as well as two dispersive bands with $E = (\omega_A + \omega_B) / 2 \pm \sqrt{(\omega_A - \omega_B)^2/4 + 2 J^2 (1 + \cos k) + g^2}$.  For the flat band, the normalised eigenstates can be calculated using the ansatz $\psi^\dag_k = \cos \theta_k b^\dag_k + e^{i \phi_k} \sin \theta_k c^\dag_k$ from which we find that $\phi_k = k/2$ and $\tan \theta_k = - (g / 2 J \cos(k/2))^{-1}$.  The Wannier coefficients are given by
\begin{align}
    w_B (r) & = \frac{1}{\sqrt{2 \pi}} \int_{-\pi}^\pi \dd k \frac{e^{i k r}}{\sqrt{1 + 4 J^2 \cos^2 (k/2) / g^2}} \\
    w_C (r) & = \frac{1}{\sqrt{2 \pi}} \int_{-\pi}^\pi \dd k \frac{e^{i k r} e^{i k / 2}}{\sqrt{1 + g^2 / 4 J^2 \cos^2 (k / 2)}}
\end{align}
Therefore, as with the sawtooth lattice, the creation operators of the Lieb lattice can be expressed as
\begin{align}
    \label{eq:b_Wannier_transform}
    b^\dag_i & = \sum_j w_B(r_i - r_j) W_j^\dag + \mathrm{higher\ bands} \\
    \label{eq:c_Wannier_transform}
    c^\dag_i & = \sum_j w_C(r_i - r_j) W_j^\dag + \mathrm{higher\ bands}
\end{align}
These are eigenstates of the Lieb Hamiltonian Eq.~(\ref{eq:HLieb_sitebasis}) so projecting onto the flat band we have $H = \sum_j \omega_B W^\dag_j W_j$.  If we introduce local, independent dissipation on the B and C sublattice, we obtain a similar non-local dissipator to the one used for the sawtooth lattice.  The dissipation rate of the A sublattice is irrelevant as the flat band eigenstates have no weight of the A sites.  

The derivation of the non-local dissipation coefficients follows from Eqs.~(\ref{eq:b_Wannier_transform}) and~(\ref{eq:c_Wannier_transform}) in exactly the same way as for the sawtooth lattice.  Therefore,
\begin{equation}
    \frac{\gamma_j}{\gamma_C} = \sum_{i} \kappa' w_B(r_{i}) w_B (r_{i} - r_{j}) + w_C(r_{i}) w_C (r_{i} - r_{j}))
\end{equation}
where $\gamma_B$ and $\gamma_C$ are the dissipation rates on the B and C sites respectively and $\kappa' = \gamma_B / \gamma_C$.  We can evaluate the terms in this sum
\begin{widetext}
\begin{align}
    \sum_{i} w_B(r_{i}) w_B (r_{i} - r_{j}) 
    & = \frac{1}{2 \pi} \int_{-\pi}^\pi \int_{-\pi}^\pi \frac{\sum_i e^{i (k + k') r_i} e^{- i k' r_j} }{\sqrt{1 + (2 J \cos (k / 2) / g)^2} \sqrt{1 + (2 J \cos (k' / 2) / g)^2}} \dd k \dd k' \\
    & = \frac{1}{2 \pi} \int_{-\pi}^\pi \frac{\cos (k j)}{1 + (2 J \cos (k/2) / g)^2} \dd k \\
    & = {}_3 \tilde{F}_2 (1/2,1,1; 1+j, 1-j; - a)
\end{align}
and
\begin{align}
    \sum_{i} w_C(r_{i}) w_C (r_{i} - r_{j}) 
    & = \frac{1}{2 \pi} \int_{-\pi}^\pi \int_{-\pi}^\pi \frac{\sum_i e^{i (k + k') r_i} e^{- i k' r_j} e^{i (k+k') / 2} }{\sqrt{1 + (g / 2 J \cos (k / 2))^2} \sqrt{1 + (g / 2 J \cos (k' / 2))^2}} \dd k \dd k' \\
    & = \frac{1}{2 \pi} \int_{-\pi}^\pi \frac{\cos (k j)}{1 + (g / 2 J \cos (k/2))^2} \dd k \\
    & = \frac{a}{2} {}_3 \tilde{F}_2 (1,3/2,2;2+j,2-j;- a)
\end{align}
where $a = (2J/g)^2$.  We can express the non-local dissipation coefficients in a similar form as for the sawtooth lattice by noting that
\begin{align}
    \frac{a}{2} {}_3 \tilde{F}_2 (1,3/2,2;2+j,2-j;- a) & = \frac{a}{2 \Gamma (2+j) \Gamma (2-j)} \sum_{k=0}^\infty \frac{(1)_k (3/2)_k (2)_k}{(2+j)_k (2-j)_k} \frac{a^k}{k!} \\
    & = \frac{-1}{2 \Gamma (2+j) \Gamma (2-j)} \sum_{k=1}^\infty \frac{(1)_{k-1} (3/2)_{k-1} (2)_{k-1}}{(2+j)_{k-1} (2-j)_{k-1}} \frac{a^k}{(k-1)!} \\
    & = \frac{-(1+j) (1-j)}{2 \Gamma (2+j) \Gamma (2-j)} \sum_{k=1}^\infty \frac{(1)_k (1/2)_k (1)_k}{(1+j)_k (1-j)_k} \frac{a^k}{k!} \\
    & = \frac{-1}{\Gamma (1+j) \Gamma (1-j)} \left(\sum_{k=0}^\infty \frac{(1)_k (1/2)_k (1)_k}{(1+j)_k (1-j)_k} \frac{a^k}{k!} - 1 \right) \\
    & =\delta_{j,0}  - {}_3 \tilde{F}_2 (1/2,1,1; 1+j, 1-j; - a) 
\end{align}
\end{widetext}
where $(x)_k$ is the Pochhammer symbol and we have used the identity $(x)_k = \Gamma(x+k)/\Gamma(x)$, which implies that $(x+1)_{k-1} = (x)_k / x$ and $(1)_{k-1} = (1)_k / k$.  Additionally, we obtain the delta-function $1 / \Gamma(1+j) \Gamma(1-j) \to \delta_{j,0}$ as $j$ takes only integer values, $\Gamma(1) = 1$ and $\modulus{\Gamma (1 - \modulus{j})} \to \infty$ for all $j \neq 0$.  Therefore
\begin{equation}
    \frac{\gamma_j}{\gamma_C} = \delta_{j,0} - (1 - \kappa) f_j \left((2 J / g)^2\right)
\end{equation}
where the geometric factor $f_j (a) =  {}_3 \tilde{F}_2 (1/2,1,1; 1+j, 1-j; -a)$ is now a function of the tunnelling rates $J$ and $g$.  

\begin{figure}
    \centering
    \includegraphics[width=\columnwidth]{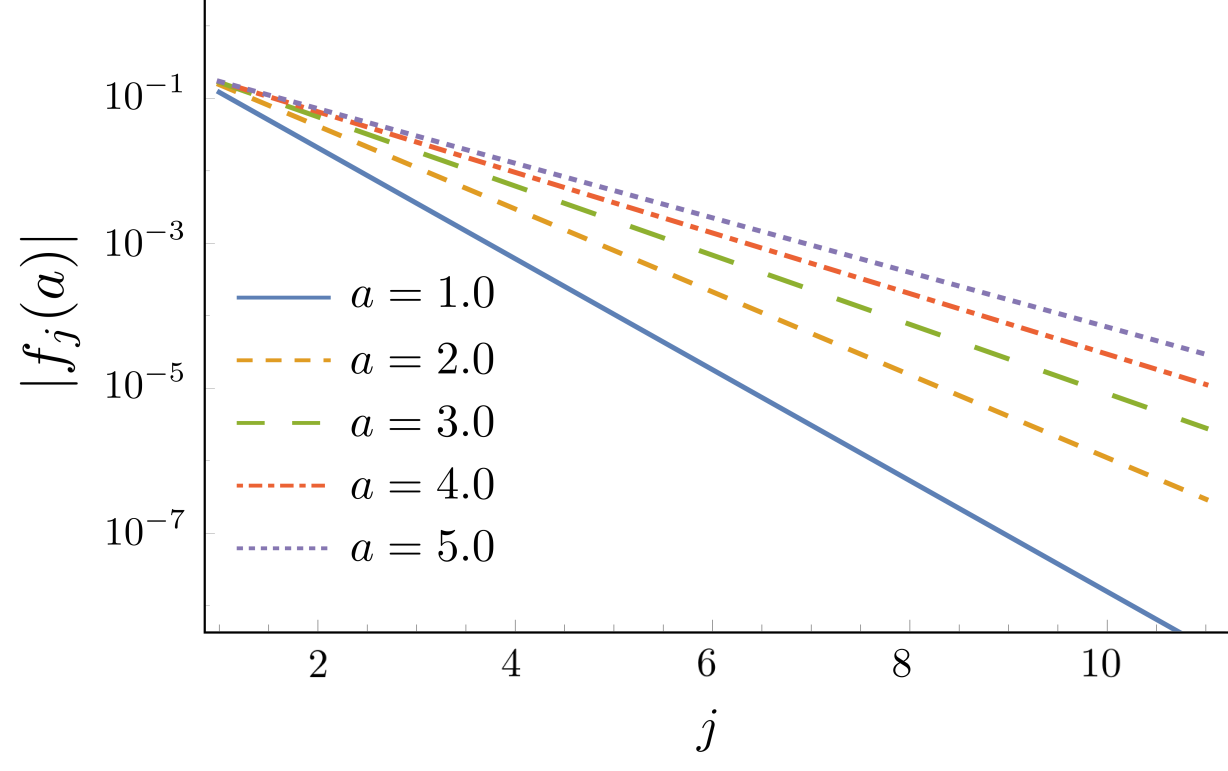}
    \caption{(Color online) The absolute value of the non-local dissipation coefficients $\modulus{f_j (a)}$ where $a = (2 J / g)^2$ is the proportional to the ratio of the couplings of the A sublattice to the B and the C sublattices.  For $a = 2.0$, the couplings are identical to the results from the sawtooth lattice.}
    \label{fig:LiebCoupling}
\end{figure}

Fig.~\ref{fig:LiebCoupling} shows the distance dependence of the magnitude of the non-local dissipation rate for a range of values of $a = (2 J / g)^2$.  We see that as $J / g$ increases, the characteristic length scale of the incoherent transport increases.  This can be understood intuitively as this coupling increases the kinetic energy along the undecorated lattice (i.e. for the system where the coupling to the C sublattice vanishes) so excitations can more easily flow along the lattice, even through dissipative processes.

\section{Derivation of the Wannier state master equation}
\label{app:MasterEquationDerivation}

In this appendix, we give a derivation of the Lindblad master equation Eq.~(\ref{eq:site_master_equation}) in the Wannier basis from an initial, general system-environment Hamiltonian and show that it is valid in the limit of weak system-environment coupling.

Initially, let us assume that there is no driving potential $\Omega = \omega_D = 0$.  We assume that the system consists of the original sawtooth lattice where each site is \emph{independently} coupled to a large set of harmonic oscillators which act as independent baths for each of the lattice sites.  The total Hamiltonian for this setup is given by
\begin{equation}
    H = H_S + H_E + H_I
\end{equation}
with
\begin{align}
    H_S & = \sum_i \omega_0 ( a^\dag_i a_i + b_i^\dag b_i ) \nonumber \\
          & \qquad + \sum_i \left(t b^\dag_{i-1} b_i + t' (b^\dag_i a_i + b^\dag_i a_{i-1}) + \mathrm{h.c.} \right) \\
    H_E & = \sum_{k, i} \omega^a_{k,i} (c^a_{k,i})^\dag c^a_{k,i} + \omega^b_{k,i} (c^b_{k,i})^\dag c^b_{k,i} \\
    H_I & = \sum_{k,i} g_{k,i}^a a_i^\dag c_{k,i}^a + g_{k,i}^b b_i^\dag c_{k,i}^b + \mathrm{h.c.}
\end{align} 
where $g_{k,i}^a$ is the coupling strength of the $i$th A site to the harmonic mode $k$, $c_{k,i}^a$ is the creation operator for this mode and $\omega_{k,i}^a$ is the mode energy and respectively for the B sites.

To find the master equation for the Wannier states, we transform to the Wannier basis using Eq.~(\ref{eq:a_operator_def}) and trace out the environmental degrees of freedom neglecting the dispersive band. For the Wannier states, $H_S$ and $H_I$ are defined as
\begin{align}
    H_S & = \sum_i \Delta W_i^\dag W_i \\
    H_I & = \sum_{i, j, k} g_{k,i}^a w_A^{ij} W_j^\dag c_{k,i}^a + g_{k,i}^b w_B^{ij} W_j^\dag c_{k,i}^b + \mathrm{h.c.}
\end{align}
where, for clarity, we have abbreviated the Wannier state coefficients $w_A (r_i - r_j) = w_A^{ij}$ and $w_B (r_i - r_j) = w_B^{ij}$.

To calculate the master equation in the Born-Markov approximation, we start from~\cite{Breuer:2002}
\begin{align}
    \label{eq:NakajimaZwanzig}
    \dot{\rho} (t) & = - i [H_S, \rho (t)] \nonumber \\
    & \qquad - \int_0^t \dd s \, \Tr_E \{ [ H_I, [ H_I (s - t), \rho_E \otimes \rho (t)] ] \}
\end{align}
where $H_I (t) = e^{i (H_S + H_E) t} H_I e^{- i (H_S + H_E) t}$ is the interaction Hamiltonian in the interaction picture, $\rho (t)$ and $\rho_E$ is the reduced density matrix of the system and environment respectively and $\Tr_E \{ \cdot \}$ is the trace over the environmental degrees of freedom.  As both $H_S$ and $H_E$ are diagonal, it is simple to show that
\begin{align}
    H_I (t) & = \sum_{i, j, k} g_{k,i}^a w_A^{ij} e^{i (\Delta - \omega_{k,i}^a) t} W_j^\dag c_{k,i}^a \nonumber \\
    \label{eq:ham-SEI}
    & \qquad + g_{k,i}^b w_B^{ij} e^{i (\Delta - \omega_{k,i}^b) t} W_j^\dag c_{k,i}^b + \mathrm{h.c.}
\end{align}
The dispersive band couples similarly to the bath through a separate interaction Hamiltonian 
\begin{align}
    H^V_I (t) & = \sum_{i, q, k} g_{k,i}^a v_A^{iq} e^{i (\Delta'_q - \omega_{k,i}^a) t} V_q^\dag c_{k,i}^a \nonumber \\
    & \qquad + g_{k,i}^b v_B^{iq} e^{i (\Delta'_q - \omega_{k,i}^b) t} V_q^\dag c_{k,i}^b + \mathrm{h.c.}
\end{align}
where $\Delta'_q$, $V_q$ and $v_{A (B)}^{iq}$ are the energies,  annihilation operators and matrix coefficients for the Bloch states of the dispersive band respectively.  By virtue of the orthogonality relations of the bath operators in Eq.~\ref{eq:bathcomm1}--\ref{eq:bathcomm3}, the terms in Eq.~\ref{eq:NakajimaZwanzig} which couple the flat and dispersive bands by including both $H_I$ and $H_I^V$ only contain contributions where the flat and the dispersive band couple to the same bath mode. Since the energy difference between the flat and dispersive band largely exceeds the strength of the system bath couplings, these terms can be neglected in a rotating wave approximation. As a consequence, it is a good approximation to restrict the description to the flat band interacting with the environments as we do in our approach. This justifies our assumption that the dispersive band does not affect the mobility of excitations in the flat band.

Evaluating the trace for each term of the double commutator is tedious so here we just focus on the first term
\begin{widetext}
\begin{align}
    & \int_0^t \dd s \, \Tr_E \{ H_I \, H_I (s - t) \rho_E \otimes \rho (t) \} \nonumber \\
    & \quad = \sum_{i,j,k} \sum_{i',j',k'} \int_0^t \dd s \, \Tr_E \Big\{ \Big( g_{k,i}^a w_A^{ij} W_j^\dag c_{k,i}^a + g_{k,i}^b w_B^{ij} W_j^\dag c_{k,i}^b + (g_{k,i}^a)^* w_A^{ij} W_j (c_{k,i}^a)^\dag + (g_{k,i}^b)^* w_B^{ij} W_j (c_{k,i}^b)^\dag \Big) \nonumber \\
    & \qquad \cdot \Big( g_{k',i'}^a w_A^{i'j'} e^{i (\Delta - \omega_{k',i'}^a) (s - t)} W_{j'}^\dag c_{k',i'}^a + g_{k',i'}^b w_B^{i'j'} e^{i (\Delta - \omega_{k',i'}^b) (s - t)} W_{j'}^\dag c_{k',i'}^b \nonumber \\
    & \qquad \qquad + (g_{k',i'}^a)^* w_A^{i'j'} e^{- i (\Delta - \omega_{k',i'}^a) (s - t)} W_{j'} (c_{k',i'}^a)^\dag + (g_{k',i'}^b)^* w_B^{i'j'} e^{-i (\Delta - \omega_{k',i'}^b) (s - t)} W_{j'} (c_{k',i'}^b)^\dag \Big) \nonumber \\
    & \qquad \cdot \rho_E \otimes \rho (t) \Big\} \\
    & \quad = \sum_{i,j,j',k} \modulus{g_{k,i}^a}^2 w_A^{ij} w_A^{ij'} \int_0^t \dd s \Big(e^{-i (\Delta - \omega_{k,i}^a) (s - t)} (n^a_{k,i} + 1) W_j^\dag W_{j'} \rho (t) + e^{i (\Delta - \omega_{k,i}^a) (s - t)} n^a_{k,i} W_j W_{j'}^\dag \rho (t) \Big) \nonumber \\
    & \qquad \qquad + \modulus{g_{k,i}^b}^2 w_B^{ij} w_B^{ij'} \int_0^t \dd s \Big(e^{-i (\Delta - \omega_{k,i}^b) (s - t)} (n^b_{k,i} + 1) W_j^\dag W_{j'} \rho (t) + e^{i (\Delta - \omega_{k,i}^b) (s - t)} n^b_{k,i} W_j W_{j'}^\dag \rho (t) \Big) \\
    & \quad = \sum_{j,j'} F_{jj'}^1 W_j^\dag W_{j'} \rho (t) + F_{jj'}^2 W_j W_{j'}^\dag \rho (t)
\end{align}
where the trace over the environmental harmonic modes, assuming that they are in thermal states, allows us to use the orthogonality relations
\begin{equation}
    \label{eq:bathcomm1}
    \Tr_E \left\{ c_{k,i}^a c_{k',i'}^a \rho_E \right\} =\Tr_E \left\{ c_{k,i}^b c_{k',i'}^b \rho_E \right\} = \Tr_E \left\{ c_{k,i}^a c_{k',i'}^b \rho_E \right\} = \Tr_E \left\{ (c_{k,i}^a)^\dag c_{k',i'}^b \rho_E \right\} = 0
\end{equation}
\begin{equation}
    \Tr_E \left\{ (c_{k,i}^a)^\dag c_{k',i'}^a \rho_E \right\} = \delta_{i, i'} \delta_{k,k'} n^a_{k,i}
\end{equation}
\begin{equation}
    \label{eq:bathcomm3}
    \Tr_E \left\{ (c_{k,i}^b)^\dag c_{k',i'}^b \rho_E \right\} = \delta_{i, i'} \delta_{k,k'} n^b_{k,i}
\end{equation}
Here $n_{k,i}^a$ is the occupation of the $k$th mode of the bath of the harmonic oscillator coupled to the A site of lattice site $i$ and similarly for the B sites and we have defined
\begin{align}
    F_{jj'}^1 & = \sum_{i,k} \int_0^t \dd s \left( \modulus{g_{k,i}^a}^2 w_A^{ij} w_A^{ij'} e^{-i (\Delta - \omega_{k,i}^a) (s - t)} (n^a_{k,i} + 1) + \modulus{g_{k,i}^b}^2 w_B^{ij} w_B^{ij'} e^{-i (\Delta - \omega_{k,i}^b) (s - t)} (n^b_{k,i} + 1) \right) \\
    F_{jj'}^2 & = \sum_{i,k} \int_0^t \dd s \left( \modulus{g_{k,i}^a}^2 w_A^{ij} w_A^{ij'} e^{i (\Delta - \omega_{k,i}^a) (s - t)} n^a_{k,i} + \modulus{g_{k,i}^b}^2 w_B^{ij} w_B^{ij'} e^{i (\Delta - \omega_{k,i}^b) (s - t)} n^b_{k,i} \right)
\end{align}
The resulting double commutator is given by
\begin{equation}
\begin{split}
	    \int_0^t \dd s \, \Tr_E \{ [ H_I, [ H_I (s - t), \rho_E \otimes \rho (t)] ] \} = \sum_{j,j'} \left\{
    F_{jj'}^1 W_j^\dag W_{j'} \rho (t) + F_{jj'}^2 W_j W_{j'}^\dag \rho (t) - F_{jj'}^2 W_j^\dag  \rho (t) W_{j'} - F_{jj'}^1 W_j \rho (t) W_{j'}^\dag \right. \\
    \left. - (F_{jj'}^2)^* W_j^\dag \rho (t) W_{j'} - (F_{jj'}^1)^* W_j \rho (t) W_{j'}^\dag + (F_{jj'}^2)^* \rho (t) W_j W_{j'}^\dag + (F_{jj'}^1)^* \rho (t) W_j^\dag W_{j'} \right\}
\end{split}
\end{equation}
\end{widetext}
Separating $F_{jj'}^1$ and $F_{jj'}^2$ into real and imaginary parts allows us to complete the master equation in the Wannier basis
\begin{align}
    \label{eq:master_LambShift}
    \dot{\rho} (t) = & - i [H_S + H_{LS}, \rho(t)] \nonumber \\
                     & +\sum_{j,j'} \frac{\gamma_{jj'}}{2} D [W_j, W_{j'}^\dag] \rho + \frac{\gamma'_{jj'}}{2} D [W_j^\dag, W_{j'}] \rho
\end{align}
where $D[A, B] \rho \equiv 2 A \rho B - \{B A, \rho\}$ and
\begin{align}
    \gamma_{jj'} & = 2 \mathrm{Re} [F_{jj'}^1] \\
    \gamma'_{jj'} & = 2 \mathrm{Re} [F_{jj'}^2]
\end{align}
The master equation Eq.~(\ref{eq:master_LambShift}) also contains a Lamb shift with
\begin{equation}
    \label{eq:LambShift}
    H_{LS} = \sum_{j,j'} \mathrm{Im} [F_{jj'}^1] W_j^\dag W_{j'} + \mathrm{Im} [F_{jj'}^2] W_j W_{j'}^\dag
\end{equation}

We will now make a couple of approximations about the environment and the system's coupling to it.  Firstly, we assume that the environment is in its ground state and contains no excitations, i.e. $n_{k,i}^a \approx 0$ and $n_{k,i}^b \approx 0$ so $F_{j,j'}^2 \approx 0$. Whereas generalizations to finite temperature environments would also yield the mobility we find, the assumption of a zero temperature environment is a very good approximation for systems hosting optical or microwave photons.  Secondly, let us assume that the spectrum and the coupling strength of each of the sites to its local environment is the same.  This means that $g_{k,i}^a$, $g_{k,i}^b$, $\omega_{k,i}^a$ and $\omega_{k,i}^b$ are all independent of $i$.  However, we allow these properties to be different between the A and B sites such that $g_{k,i}^a \neq g_{k,i}^b$ and $\omega_{k,i}^a \neq \omega_{k,i}^b$ so that the dissipation rates on the A and B sites can differ.  
The non-local dissipation rates can be calculated  directly from this calculation
\begin{widetext}
\begin{align}
    \gamma_{jj'} = \left\{ \begin{array}{ll}
                    2 \int_0^t \dd s\sum_k (f_0 + f_1) \modulus{g_k^a}^2 \cos \left[ (\Delta - \omega_k^a) (s - t) \right] + f_0 \modulus{g_k^b}^2 \cos \left[ (\Delta - \omega_k^b) (s - t) \right] & \mathrm{for\ } j = j'\\
                    - 2 f_l \int_0^t \dd s \sum_k \modulus{g_k^a}^2 \cos \left[ (\Delta - \omega_k^a) (s - t) \right] - \modulus{g_k^b}^2 \cos \left[ (\Delta - \omega_k^b) (s - t) \right] & \mathrm{for\ } j - j' = l
    \end{array} \right.
\end{align}
\end{widetext}
where $f_l = (\sqrt{3} - 2)^{\modulus{l}} / \sqrt{3}$ is calculated using the residue theorem as shown in App.~\ref{app:DissipationWannier}.

An inhomogeneity in the Lamb shifts described in Eq.~(\ref{eq:LambShift}) would lead to a kinetic energy term an thus to additional mobility.  These Lamb shifts are equivalent to local frequency shifts, 
\begin{align}
    H_S' & = \sum_i \delta \omega_A a^\dag_i a_i + \delta \omega_B b^\dag_i b_i \\
         & = \sum_{i, j, j'} (\delta \omega_A w_A^{ij} w_A^{ij'} + \delta \omega_B w_B^{ij} w_B^{ij'} ) W^\dag_j W_{j'}
\end{align}
and are therefore irrelevant for most experimental situations. 

In photonic waveguide lattices, the propagation distance along the waveguide takes on the role of time and the propagation constant the role of the resonance frequency. The latter is set by the real part of the refractive index which is unaffected by the loss channels that are patterned into the chip via additional waveguides. For lattices of coupled resonators, in turn, one aims to build mutually resonant lattice sites. Therefore the lattice sites with larger damping rates should be built to have a correspondingly adjusted frequency, which is not more difficult than building them at exactly equal frequencies. Any residual frequency disorder can in many experimental situations be compensated by tuning the individual lattice sites, e.g. via ac-Stark shifts. Since any measurement of the resonance frequency of a resonator always includes the Lamb shift, disorder in these shifts is then automatically compensated for when tuning the device.

\section{Decay Rate Derivation for the Effective Drive Model}
\label{app:DecayRateDerivation}

In this appendix, we derive the decay length of the density profile for the non-interacting system using an effective drive model.  This model involves solving the equations of motion for the three site system $i = \{0, j, j+1\}$ where $j > 1$ assuming that the two unpumped sites do not affect the field amplitude of site $i = 0$, which is a constant determined by the three site system $i = \{-1, 0, 1\}$.  This model combines nearest-neighbour and direct non-local dissipative coupling terms and results in an accurate prediction of the decay length $\xi$ for a large range of dissipation rate asymmetry $\kappa$.

To calculate the field amplitude for the pumped site, we start by solving Eq.~(\ref{eq:Field_Expectation}) for the system $i = \{-1, 0, 1\}$.  The equations of motion are given by
\begin{align}
    \label{eq:effective_drive_pumped_site_amplitude}
    \frac{d \expectation{W_0}}{d t} & \approx - \frac{i}{2} \Omega_W^* - \gamma_0 \expectation{W_0} - 2 \gamma_1 \expectation{W_1} \\
    \label{eq:effective_drive_nearest_neighbour_amplitude}
    \frac{d \expectation{W_1}}{d t} & \approx - \gamma_0 \expectation{W_1} - \gamma_1 \expectation{W_0}
\end{align}
where we have exploited the parity symmetry around $i = 0$ and the relation $\gamma_l = \gamma_{-l}$.  Solving Eqs.~(\ref{eq:effective_drive_pumped_site_amplitude}) and~(\ref{eq:effective_drive_nearest_neighbour_amplitude}) for the pumped site's steady state amplitude gives
\begin{equation}
    \expectation{W_0^\dag} \approx \frac{-i \gamma_0 \Omega_W^*}{2 (\gamma_0^2 - 2 \gamma_1^2)}
\end{equation}
which we can insert into the steady state of Eq.~(\ref{eq:Field_Expectation}) along with the assumption that $\sum_{l \neq \{0, 1\} } \gamma_l \expectation{W_{j + l}} \approx \gamma_{-j} \expectation{W_0}$ which we made in Sec.~\ref{subsec:EffectiveDrive}
\begin{align}
    \label{eq:eff_drive_eq1}
    0 & \approx \frac{- i \gamma_0 \gamma_j \Omega_W^*}{2 (\gamma_0^2 - 2 \gamma_1^2)} - \gamma_0 \expectation{W_j} - \gamma_1 \expectation{W_{j+1}} \\
    \label{eq:eff_drive_eq2}
    0 & \approx \frac{- i \gamma_0 \gamma_{j+1} \Omega_W^*}{2 (\gamma_0^2 - 2 \gamma_1^2)} - \gamma_0 \expectation{W_{j+1}} - \gamma_1 \expectation{W_j}
\end{align}
We see that the pumped site acts as an effective drive for the two site system with
\begin{align}
    \Omega_{W, j}^* &= \frac{\gamma_0 \gamma_j \Omega_W^*}{\gamma_0^2 - 2 \gamma_1^2} \\
    \Omega_{W, j+1}^* &= \frac{\gamma_0 \gamma_{j+1} \Omega_W^*}{\gamma_0^2 - 2 \gamma_1^2}
\end{align}
These effective drives can be used in the equations of motion for the correlations of the Wannier state operators in the steady state, namely
\begin{widetext}
\begin{align}
    \label{eq:eff_drive_eq3}
    \gamma_0 \langle W^\dag_j W_j \rangle & \approx \frac{i}{2} \left(\Omega_{W, j} \langle W_j \rangle - \Omega^*_{W, j} \langle W^\dag_j \rangle \right) - \frac{\gamma_1}{2} \left( \langle W^\dag_j W_{j+1} \rangle + \langle W^\dag_{j+1} W_j \rangle \right) \\
    \gamma_0 \langle W^\dag_j W_{j+1} \rangle & \approx \frac{i}{2} \left(\Omega_{W, j} \langle W_{j+1} \rangle - \Omega^*_{W, j+1} \langle W^\dag_j \rangle \right) - \frac{\gamma_1}{2} \left( \langle W^\dag_j W_j \rangle + \langle W^\dag_{j+1} W_{j+1} \rangle \right) \\
    \gamma_0 \langle W^\dag_{j+1} W_j \rangle & \approx \frac{i}{2} \left(\Omega_{W, j+1} \langle W_j \rangle - \Omega^*_{W, j} \langle W^\dag_{j+1} \rangle \right) - \frac{\gamma_1}{2} \left( \langle W^\dag_j W_j \rangle + \langle W^\dag_{j+1} W_{j+1} \rangle \right) \\
    \label{eq:eff_drive_eq6}
    \gamma_0 \langle W^\dag_{j+1} W_{j+1} \rangle & \approx \frac{i}{2} \left(\Omega_{W, j+1} \langle W_{j+1} \rangle - \Omega^*_{W, j+1} \langle W^\dag_{j+1} \rangle \right) - \frac{\gamma_1}{2} \left( \langle W^\dag_j W_{j+1} \rangle + \langle W^\dag_{j+1} W_j \rangle \right) 
\end{align}
\end{widetext}
We can solve Eqs.~(\ref{eq:eff_drive_eq1})--(\ref{eq:eff_drive_eq6}) to find approximate densities on the sites $j$ and $j+1$ from which we can calculate the decay length
\begin{equation}
    \xi \approx \left[2 \log_{10} \left(\frac{\gamma_0 \gamma_j - \gamma_1 \gamma_{j+1}}{\gamma_1 \gamma_j - \gamma_0 \gamma_{j+1}} \right)\right]^{-1}
\end{equation}
As discussed in Sec.~\ref{subsec:EffectiveDrive}, this provides a good approximation for the exact density decay length $\xi$ for a large range of $\kappa$.

\section{Interactions in the Wannier Basis}

\subsection{Calculating the Interaction Strength Coefficients}
\label{app:IntCoeffCalc}

Due to the delocalized nature of the Wannier states, onsite interactions in site basis lead to non-local quartic operators in the Wannier basis.  In this appendix, we derive the coefficients of these non-local interactions in their integral form and tabulate the values.

We introduce interactions in the sawtooth lattice through a Hubbard interaction with varying interactions strengths for the A and B sublattices
\begin{equation}
    \label{eq:H_U}
    H_U = \sum_n \left( U_A a^\dag_n a^\dag_n a_n a_n + U_B b^\dag_n b^\dag_n b_n b_n \right)
\end{equation}
We now need to substitute the site operators with their Wannier state representations given in Eq.~(\ref{eq:a_operator_def}). 
\begin{widetext}
\begin{align}
    U_A \sum_n a^\dag_n a^\dag_n a_n a_n = & U_A \sum_{i, j, l, m, n} w_A (r_n - r_i) w_A (r_n - r_j) w_A (r_n - r_l) w_A (r_n - r_m) W^\dag_i W^\dag_j W_l W_m \\
            & = \frac{4 U_A}{(2 \pi)^4} \sum_{i, j, l, m} \int_{-\pi}^\pi \int_{-\pi}^\pi \int_{-\pi}^\pi \int_{-\pi}^\pi \dd k \dd k' \dd q \dd q' \nonumber \\
            & \qquad \qquad \qquad \qquad \qquad \qquad \qquad \cdot \frac{ \cos (k / 2) \cos (k' / 2) \cos (q / 2) \cos (q' / 2) \sum_n e^{-i r_n (k + k' + q + q')}}{\sqrt{(\cos k + 2)(\cos k' + 2)(\cos q + 2)(\cos q' + 2)}} \nonumber \\
            & \qquad \qquad \qquad \qquad \qquad \qquad \qquad \cdot e^{ i k r_i } e^{ i k' r_j } e^{ i q r_l }e^{ i q' r_m } e^{i (k + k' + q + q') / 2} W^\dag_i W^\dag_j W_l W_m \\
            & = \frac{4 U_A}{(2 \pi)^3} \sum_{i, j, l, m} \int_{-\pi}^\pi \int_{-\pi}^\pi \int_{-\pi}^\pi \dd k' \dd q \dd q' \frac{ \cos ((k' + q + q') / 2) \cos (k' / 2) \cos (q / 2) \cos (q' / 2)}{\sqrt{(\cos (k' + q + q') + 2)(\cos k' + 2)(\cos q + 2)(\cos q' + 2)}} \nonumber \\
            & \qquad \qquad \qquad \qquad  \qquad \qquad \qquad \cdot e^{ i k' (r_j - r_i) } e^{ i q (r_l - r_i) }e^{ i q' (r_m - r_i) } \Pi \left(\frac{(k' + q + q')}{2 \pi}\right) \nonumber \\
            & \qquad \qquad \qquad \qquad  \qquad \qquad \qquad \cdot W^\dag_i W^\dag_j W_l W_m \\
            \label{eq:UA_Wannier}
            & = \sum_{i} \sum_{j', l', m'} U^{\mathrm{eff}, A}_{j', k', l'} W^\dag_i W^\dag_{i+j'} W_{i+l'} W_{i+m'}
\end{align}
and for the B sites
\begin{align}
    U_B \sum_n b^\dag_n b^\dag_n b_n b_n
            & = U_B \sum_{i, j, l, m, n} w_B (r_n - r_i) w_B (r_n - r_j) w_B (r_n - r_l) w_B (r_n - r_m) W^\dag_i W^\dag_j W_l W_m \\
            & = \frac{U_B}{(2 \pi)^4} \sum_{i, j, l, m} \int_{-\pi}^\pi \int_{-\pi}^\pi \int_{-\pi}^\pi \int_{-\pi}^\pi \dd k \dd k' \dd q \dd q' \frac{e^{ i k r_i } e^{ i k' r_j } e^{ i q r_l }e^{ i q' r_m } \sum_n e^{-i r_n (k + k' + q + q')}}{\sqrt{(\cos k + 2)(\cos k' + 2)(\cos q + 2)(\cos q' + 2)}} \nonumber \\
            & \qquad \qquad \qquad \qquad \qquad \qquad \qquad \qquad \qquad \cdot W^\dag_i W^\dag_j W_l W_m \\
            & = \frac{U_B}{(2 \pi)^3} \sum_{i, j, l, m} \int_{-\pi}^\pi \int_{-\pi}^\pi \int_{-\pi}^\pi \dd k' \dd q \dd q' \frac{ e^{ i k' (r_j - r_i) } e^{ i q (r_l - r_i)}e^{ i q' (r_m - r_i) }}{\sqrt{(\cos (k' + q + q') + 2)(\cos k' + 2)(\cos q + 2)(\cos q' + 2)}} \nonumber \\
            & \qquad \qquad \qquad \qquad \qquad \qquad \qquad \qquad \qquad \cdot \Pi \left(\frac{(k' + q + q')}{2 \pi}\right) W^\dag_i W^\dag_j W_l W_m \\
            \label{eq:UB_Wannier}
            & = \sum_{i} \sum_{j', l', m'} U^{\mathrm{eff}, B}_{j', l', m'} W^\dag_i W^\dag_{i+j'} W_{i+l'} W_{i+m'}
\end{align}
where
\begin{align}
    U^{\mathrm{eff}, A}_{j', l', m'} & = \frac{4 U_A}{(2 \pi)^3} \int_{-\pi}^\pi \int_{-\pi}^\pi \int_{-\pi}^\pi \dd k' \dd q \dd q' \frac{\cos ((k' + q + q') / 2) \cos (k' / 2) \cos (q / 2) \cos (q' / 2)}{\sqrt{(\cos (k' + q + q') + 2)(\cos k' + 2)(\cos q + 2)(\cos q' + 2)}} \nonumber \\
            & \qquad
            \qquad \qquad \qquad \qquad \qquad \qquad \qquad \qquad \qquad \cdot e^{i k' j'} e^{i q l'} e^{i q' m'} \cdot \Pi \left(\frac{(k' + q + q')}{2 \pi}\right)  \\
    U^{\mathrm{eff}, B}_{j', l', m'} & = \frac{U_B}{(2 \pi)^3} \int_{-\pi}^\pi \int_{-\pi}^\pi \int_{-\pi}^\pi \dd k' \dd q \dd q' \frac{e^{i k' j'} e^{i q l'} e^{i q' m'} }{\sqrt{(\cos (k' + q + q') + 2)(\cos k' + 2)(\cos q + 2)(\cos q' + 2)}} \cdot \Pi \left(\frac{(k' + q + q')}{2 \pi}\right) 
\end{align}
\end{widetext}
are the effective interaction strengths and $\Pi (x)$ is the top-hat function of unit width centered at zero. The  $U^{\mathrm{eff}, A}_{j', l', m'}$ and $U^{\mathrm{eff}, B}_{j', l', m'}$ are symmetric on interchange of the indices $j', l'$ and $m'$ and $U^{\mathrm{eff}, B}_{j', l', m'}$ is also symmetric for $j' \to -j'$, $l' \to -l'$ and $m' \to -m'$.  These effective interaction strengths decrease exponentially with increasing $j', l'$ and $m'$ as they consist of sums over the Wannier coefficients, which are exponentially localised.

\subsection{Truncating to the single excitation subspace for MPO simulations}
\label{app:SingleExcitationSubspace}

The sum in Eqs.~(\ref{eq:UA_Wannier}) and~(\ref{eq:UB_Wannier}) describe an infinite number of interaction processes.  However, these interactions can be divided into three types: density-assisted tunnelling, cotunnelling and density-density interactions.  Density-assisted tunnelling terms produce coherent coupling between two sites which is activated when there is an excitation either on one of the tunnelling sites or on a neighbouring site.  An example is given by $j' = 1, l' = 1, m' = -1$ which describes coherent tunnelling between the sites $i$ and  $i - 1$ which occurs only when there is an excitation on site $i + 1$.  Cotunnelling terms describe two excitations on different lattice sites tunnelling together onto two other lattice sites.  An example would be when $j' = -1, l' = 1, m' = 2$ which corresponds to a tunnelling of an excitation on the $i$ and $i - 1$ sites to the $i + 1$ and $i + 2$ sites.  Finally, density-density interactions terms with a cross-Kerr non-linearity such as for $j' = 1, l' = 0, m' = 1$ which gives an energy penalty for having excitations on neighbouring sites.

Whilst the density-density interactions cannot transfer excitations between sites, the density-assisted tunnelling and cotunnelling terms both introduce coherent transfer processes.  We want to investigate how interactions affect the dissipation-induced mobility we examine in this paper.  We expect that the cross-Kerr non-linearity will decrease the mobility as it introduces an energy shift on sites neighbouring an excitation.  This will detune the neighbouring site and prevent excitations moving into these states due to energy conservation.  On the other hand, the density-assisted tunnelling and cotunnelling terms should increase the mobility.

This proliferation of possible transfer processes makes it difficult to determine which processes are affecting the density distribution when including all interaction terms in Eqs.~(\ref{eq:UA_Wannier}) and~(\ref{eq:UB_Wannier}).  In order to reduce the complexity of this problem, we choose to work in the strongly-interacting limit where the probability that a site contains two or more excitations is vanishingly small. This means that the operator $H_U$ vanishes when two Wannier state annihiliation or creation operators operate on the same site.  

This truncation also facilitates our calculations of the density profiles in the interacting regime using a variational Matrix Product Operator (MPO) method~\cite{Schollwock:2011, Mascarenhas:2015, Cui:2015}.  The computational resources necessary for this method decrease significantly if the system can be described by a reduced Hilbert space on each lattice site.   For our truncation approximation to be valid, we need $U_0$ to be large enough to prevent the pump doubly-exciting the site at $i = 0$.

\begin{figure}
    \centering
    \includegraphics[width=\columnwidth]{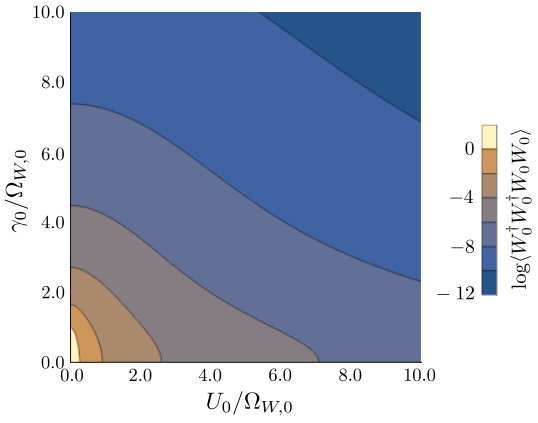}
    \caption{(Color online) Second-order correlation function for the pumped site $\langle W^\dag_0 W^\dag_0 W_0 W_0 \rangle$ (on a logarithmic scale) with coherent and incoherent transfer processes set to zero.  This provides an upper bound on the probability that a site is occupied by multiple excitations.}
    \label{fig:doubleocc_prob}
\end{figure}

To see when this approximation is valid, we calculate the second order correlation function for the pumped site $\langle W^\dag_0 W^\dag_0 W_0 W_0 \rangle$ assuming that all coherent and incoherent processes coupling it to other sites vanish.  This gives us an upper bound on the probability that the pumped site contains at least two excitations as the addition of non-local terms will only decrease the density on the pumped site.  For this case, $\langle W^\dag_0 W^\dag_0 W_0 W_0 \rangle$ can be calculated exactly using an approach by Drummond and Walls~\cite{Drummond:1980, LeBoite:2013}. The result is plotted in Fig.~\ref{fig:doubleocc_prob} as a function of $U_0 / \Omega_{W, 0}$ and $\gamma_0 / \Omega_{W, 0}$.

Using the simulations performed in the main text allows us to perform a consistency check on the truncated Hilbert-space approximation for the parameters we have used.  The excitation density of the site next the pump for the non-interacting system is observed to be in the range of $10^{-2} - 10^{-3}$ for $\kappa > 0.1$.  We want the probability of finding a double-excitation to be smaller than this.  In our simulations, we set $\gamma_A = 1$ and $\Omega_{W, 0} = 1$ so for $\langle W^\dag_0 W^\dag_0 W_0 W_0 \rangle \lesssim 10^{-4}$, Fig.~\ref{fig:doubleocc_prob} shows that we require $U_0 \gtrsim 2.0$.

This provides us with a minimum onsite interaction strength $U_0$ which we must use to guarantee the validity of the single excitation subspace approximation.  Importantly, the non-local interaction strengths where $j'$, $l'$ and $m'$ are not equal to zero are related to $U_0$ so we are unable to make them arbitrarily small.  However, we can calculate minimum value for these non-local terms which correspond to a regime where the truncation to the single excitation subspace is valid.

\subsection{Approximate Interaction Hamiltonian for the Single Excitation Subspace}

As described above, in our MPO simulations we need to truncate the Hilbert space of each lattice site to the subspace of at most one excitation.  Therefore, we exclude terms with two annihilation or creation operators on the same site, with the exception of the onsite interaction.  In this appendix, we calculate what the most relevant terms in Eqs.~(\ref{eq:UA_Wannier}) and~(\ref{eq:UB_Wannier}) are in the single excitation subspace.  We list the leading order values for $U^{\mathrm{eff}, A}_{j',l',m'}$ and $U^{\mathrm{eff}, B}_{j',l',m'}$ below:
\begin{equation}
    \begin{array}{l|c}
        \{j', l', m'\} & U^{\mathrm{eff}, A}_{j',l',m'} / U_A \\
        \hline
        \{0, 0, 0\}    & 0.028032     \\
        \{1, 1, 0\}    & 0.007568     \\
        \{2, 1, 0\}    & 0.000748     \\
        \{2, 2, 0\}    & 0.000582
    \end{array} \nonumber
\end{equation}
and
\begin{equation}
    \begin{array}{l|c}
        \{j', l', m'\} & U^{\mathrm{eff}, B}_{j',l',m'} / U_B \\
        \hline
        \{0, 0, 0\}    & 0.191937      \\
        \{1, 1, 0\}    & 0.033196      \\
        \{2, 2, 0\}    & 0.013613      \\
        \{-1, 0, 1\}   & 0.011298      \\
        \{2, 1, 0\}    & 0.009605      
    \end{array} \nonumber
\end{equation}
where $\{j', l', m'\}$ consists of a set of terms grouped using the symmetries of $U^{\mathrm{eff}, A}_{j', l', m'}$ and $U^{\mathrm{eff}, B}_{j', l', m'}$ described in the previous section of this appendix.  Terms which bring the state out of the single excitation subspace are not included.  

Including interaction effects for both A and B sublattices is an unnecessary complication, as it only leads to minor modifications in the weighting of the interaction strengths.  We have tested that these small modifications only have a negligible effect.  We choose $U_A = 0$ and $U_B \ne 0$ as this leads to the strongest possible coherent propagation terms and allows us to show that even here the incoherent mobility is the dominant effect.

In the strongly-interacting regime, the leading order terms in Eq.~(\ref{eq:H_U}) in the Wannier basis is the onsite energy with $\{j', l', m'\} = \{0, 0, 0\}$,  the nearest-neighbour cross-Kerr interaction with $\{j', l', m'\} = \{1, 1, 0\}$, the next-to-nearest-neighbour cross-Kerr interaction with $\{j', l', m'\} = \{2, 2, 0\}$ and the density-assisted tunnelling term with $\{j', l', m'\} = \{-1, 0, 1\}$.  However, the symmetries of $U^{\mathrm{eff}, B}_{j', l', m'}$ provide multiplicative factors for equivalent terms in the sum in Eq.~(\ref{eq:UB_Wannier}).  The onsite term does not have a multiplicative factor.  The cross-Kerr interactions have a fourfold multiplicity, as can be seen for density-density interactions between sites $i=0$ and $i=1$ (i.e. $W^\dag_0 W_0 W^\dag_1 W_1$) where the following terms all contribute to the interaction strength:
\begin{equation}
    \begin{array}{c|c|c|c}
        i & j' & l' & m' \\
        \hline
        0 & 1 & 0 & 1   \\
        0 & 1 & 1 & 0   \\
        1 & -1 & 0 & -1   \\
        1 & -1 & -1 & 0
    \end{array} \nonumber
\end{equation}

For the density-assisted tunnelling terms in the symmetry set $\{j', l', m'\} = \{-1, 0, 1\}$, where the density operator is on the central site and the excitation hops over it, e.g. $W^\dag_{-1} (W^\dag_0 W_0) W_1$ where $i = 0$, $j' = -1$, $l' = 0$ and $m' = 1$.  This process has a twofold multiplicity in the sum in Eq.~(\ref{eq:UB_Wannier}) due to the symmetry $j' \leftrightarrow m'$.

Considering these leading order terms, we can approximate the interaction Hamiltonian by
\begin{align}
    H_U = & \sum_i U_A a^\dag_i a^\dag_i a_i a_i + U_B b^\dag_i b^\dag_i b_i b_i \\
        \approx & \sum_i U_0 W^\dag_i W^\dag_i W_i W_i + U_1 W^\dag_i W_i W^\dag_{i+1} W_{i+1} \nonumber \\
        & \quad + U_2 W^\dag_i W_i W^\dag_{i+2} W_{i+2} \nonumber \\
        \label{eq:truncated_HU}
        & \quad + U_3 (W^\dag_{i-1} W^\dag_i W_i W_{i+1} + \mathrm{h. c.})
\end{align}
where $U_0 \approx 0.192 U_B$, $U_1 \approx 0.133 U_B$, $U_2 \approx 0.054 U_B$ and $U_3 \approx 0.023 U_B$, which is the approximation to the interaction Hamiltonian $H_U$ which we use for the MPO calculations.  Using the minimum value for $U_0$ needed to apply the single excitation subspace truncation which we calculated in the previous section of this appendix, we see that the approximation in Eq.~(\ref{eq:truncated_HU}) is valid for $U_B \gtrsim 10.0$.

\subsection{Agreement of results for interacting and noninteracting cases}
\label{app:MPODetails}

The MPO method expresses the vectorised density matrix $\bar{\rho}$ as a matrix product state (MPS) and the Lindbladian $\mathcal{L}$ as an MPO. The MPS for this system had a virtual dimension of 40 and the non-local dissipation was cut off at $l > 3$. To find the stationary state we seek the ground state of $\mathcal{L}^{\dagger}\mathcal{L}$ in the manner of \cite{Cui:2015}. 

We note that for low enough densities, our numerical MPO calculations are in agreement with the results for the noninteracting case. However, a direct comparison of interacting and noninteracting scenarios for higher densities would require the simulation of more states per lattice site which is beyond the capabilities of our MPO simulations.

\bibliographystyle{unsrt}
\bibliography{citations}

\end{document}